\documentclass[11pt]{article}
\usepackage{amsmath}
\usepackage{amsfonts}
\usepackage{epsfig}
\renewcommand{\phi}{\varphi}

\newcommand{\beq}{\begin{equation}}
\newcommand{\eeq}{\end{equation}}

\newcommand{\pd}{\partial}

\newcommand{\du}{\frac{\partial}{\partial u}}
\newcommand{\dw}{\frac{\partial}{\partial w}}
\newcommand{\ddu}{\frac{\partial^2}{\partial u^2}}
\newcommand{\ddw}{\frac{\partial^2}{\partial w^2}}

\title{\bf Probabilities in Toy Regge models with odderons}
\author{M.A. Braun\\
Dept. of High Energy physics,
Saint-Petersburg State University,\\
198504 S.Petersburg, Russia}

\begin{document}
\maketitle
\begin{abstract}
The possibi;lity of a probabilistic interpretation is studied for the Regge-Gribov model in  zero dimensional
transverse world ("Toy") with interacting pomerons and odderons. It is found that the old recipe ~\cite{redif},
which allowed to introduce such interpretation in the model without odderons, does not work once odderons are
included. Starting from the physically reasonable probabilites it leads to a pathological field theory violating
$C$-parity invariance. {\it Vice versa}, starting from an admissible field theory one comes to pathological probabilities,
which not only violate $C$-nvariance but also  allow particles to be created from or annihilated into the vacuum.
A method is proposed to introduce reasonable probabilities besed directly on  the Fock components of the wave function.
Such  probabilities manifest themselves in the multiplicity distributions of hadrons produced in high-energy
collisions. The corresponding   entropy  grows with rapidity but saturates in the limit. It is found to be similar for $C=\pm1$ processes.
\end{abstract}

\section{Introduction}

In the studies of the high-energy behavior of  amplitudes generated by strong interactions a notable place has been taken by
the Regge-Gribov models  based on the exchange of local  pomerons evolving in rapidity and
self-interacting with a non-Hermitian Hamiltonian. Unforrunateky
in the real two-dimensional transverse world they cannot be solved exactly.
For this reason much attention was given in the past to the approximation of zero slope "Toy models", in which  the models reduce to fields depending only on rapidity
("zero-dimensional" in the transverse variables). In this approximation the models actually transform to one dimensional quantum mechanical systems
living in the imaginary time $t=-iy$, where $y$ is the rapidity.
These Toy Regge-Gribov (TRGM) models well describe actual interactions with nuclei in the quasi-classical approximation, when the monenta of exchanged pomerons are small (of the
nuclear order). In contrast in the quantum version they have
a very indirect relation to realistic observables. However studies of TRGM have given valuable lessons
to be taken into account in more physical theories like QCD. In particular older results in the past century ~\cite{schwimmer,jengo,amati}
taught that loops play a decisive role in the high-energy behavior changing it radically..

Later it was proposed to look at TRGM  in the framework of the so-called statistical reaction-diffusion
approach ~\cite{redif}. In this approach the system was described by a set of probabilities $P(n)$ for existence of a given number
$n$ of particles A which transform according to the patterns A$\to$A+A and A+A$\to$A  with certain rates in accordance to what actually
 occurs in the system of pomerons. Simple reasoning then gives equations governing evolution of probabilty, with the  growth of rapidity $y$
\beq
\frac{dP(n)}{dy}\equiv\dot{P}(n)
=
\dot{P}(n)_{\alpha}+\dot{P}(n)_\beta,\label{eq3}\eeq
where the first term in the r.h.s. is the contribution of splitting: A$\to$A+A with rate $\alpha$
\beq
\dot{P}(n)_\alpha
=\alpha [(n -1)P(n - 1) - nP(n)] ,\label{eq4}\eeq
while the second one is due to recombination: A+A$\to$A with rate $2\beta$
\beq
\dot{P}(n)_\beta
=\beta[n(n + 1)P(n + 1) - n(n - 1)P(n)] . \label{eq5}\eeq
So the total evolution obeys the equation
\beq
\dot{P}(n)
=\alpha [(n -1)P(n - 1) - nP(n)]
+\beta[n(n + 1)P(n + 1) - n(n - 1)P(n)] . \label{eq5tot}\eeq

In an attempt to relate this simple probabilistic picture to TRGM in ~\cite{redif} it was proposed to relate
 the so-called multiple moments of the distribution
 \beq
 \nu(k)=\sum_{n=k}^\infty P(n)n(n-1)...(n-1+k)
 \label{nup}
 \eeq
 with the coefficients $c(k)$ in the Fock expansion of the TRGM wave function $\Psi(y)$  in terms of $n$-pomeron states
 \beq
 \Psi(y)=\sum_{n=1}^\infty c(y,k)\Psi_k.
 \label{cnu}
 \eeq
 Here $\Psi_k$ is a state with exactly $k$ bare pomerons, that is satisfying $H_0\Psi_k=-k\mu\Psi_k$ where $H_0$ is the bare
 Hamiltonian with no interaction between pomerons.
 The key proposition was the relation between $c(k)$ and $\nu(k)$
 \beq
 c(k)=-(-a)^k\frac{\nu(k)}{k!}
 \label{rel}
 \eeq
 with some constant $a$ determined by the adopted normlization.
 Relation (\ref{rel}) establishes the connection between the probabilistic and field-theoretical pictures and
 so allows to introduce probabilities and entropy into TRGM.

 The study in ~\cite{redif} discovered that this relation is not  working in the general case. In particular for the
 minimal TRGM with only triple pomeron interaction it leads to probabilities which do not agree with above mentioned evident probabilistic
 picture. They additionally introduce  annihilation $A+A$ to vacuum and some probabilities appear with negative signs. The only TRGM found to admit
 the probabilisic interpretation was the model with quartic interaction of a special type and specific coupling constant.

In the QCD, apart from the pomeron with the
positive $C$-parity and signature, a compound state of three reggeized gluons with
the negative $C$-parity and signature, the odderon, appears. Its possible experimental
manifestations have not been found with certainty up to now, which may be
explained both by its behaviour with energy and its small coupling
to participant hadrons. On the theoretical level
the QCD odderon has been extensively discussed as in its formation
 ~\cite{bar,jar,kwi,gln,lip,fad,jw1,jw2,blv,ewerz} and also in its interaction with the QCD pomeron
 in no-loop approximation ~\cite{ksw, hiim}.

Learning the lesson from the studies of the theory
with only pomerons,  in \cite{bkv2021} we proposed a generalized toy Regge-Gribov model (GTRGM)
to include interactions of both
pomerons and odderons. Using numerical technique we could find the behaviour of the propagators
and amplitudes in GTRGM. Naturally the question arises if
this generalized model admits a probabilistic interpretaion so that one can introduce  propabilities $P(n,m)$ in GTRGM to
find $n$ pomerons and $m$ odderons.  Our answer to this question is negative, provided one connects the field theory
and probabilistic picture by the relation trivially generalizing (\ref{rel}). Starting from the probabilistic approach one arrives at
the corresponding GTRGM with inappropriate Hamiltonian, containing direct transformation of the pomeron into
odderon in contradiction with their signatures. On the other hand, starting from the adequate GTRGM one arrives at the
inappropriate probabilities incompatible with the probabilistic approach to the
pomeron-odderon system.

It remains an open question if  both pictures, probabilistic and field theoretical, can be related in a manner
fundamentally differeint from (\ref{rel}) once odderons appear in the system. We leave the search for such a novel relation
for future studies.

At present we propose a method to construct in GTRM valid probabilities derived directly from Fock's coefficients of the
wave function after certain rescaling which genralizes the procedure introduced in ~\cite{braun2025} for the model without odderons.

Ihe paper is organized as follows. In Sections 2 and 3 we consider the pomeron-odderon system
in the purely probabilistic formulation and its reflection in the field theoretical framework  which
follows from certain evidemt generalization of (\ref{rel}). In Sections 4 and 5 inversely
we consider the Toy model and obtain probabilities from it with the same recipe.  In Section 6 we formulate the rescaling method to construct probabilities
in GTRGM and present some numerical results for thus constructed probabilities and entropy. In Cnclusion we make the final verdict
that contrary to the model without odderons inclusion of the latters spoils the picture altogether and does not leave  a single case
when  the probabilstic and field-theoretical pictures are compatible. Also some coclusions are extracted feom the numerical results from our rescaling method.

\section{Pomerons plus odderons. Probabilistic viewpoint}
\subsection{Probabilities}
We assume that there exist two types of particles A and B (pomerons and odderons) with the probability $P(y.n,m)$ for
$n$ particles A and $m$ particles B at rapidity $y$. In the following  argument $y$ will be often suppressed, since the relations will
refer to  fixed $y$. Note that $P(n,m)$ are defined for non-negative $n$ and $m$ with $n+m>0$, so that only $P(0,0)$ is not defined.
Particle  transitions obey the physically obvious conservation laws and are splittings
\beq
A\to A+A\ (\alpha_1),\ \ A\to B+B\ (\alpha_2),\ \ B\to A+B\ (\alpha_3)
\label{eq21}
\eeq
and fusions
\beq
A+A\to A\ (2\beta_1).\ \ B+B\to A\ (2\beta_2).\ \ A+B\to B\ (\beta_3),
\label{eq22}
\eeq
with the corresponding rates shown in brackets.
Let us see the evolution in rapidity of  probabilities $P(y,n,m)$  due to these
transitions.

Transition $A\to A+A$. Here $(n-1,m)$ particles A,B (probability $P(n-1,m)$) go into $(n,m)$ particles $A$ (probability $P(n,m)$
with the total rate
$(n-1)\alpha_1$ and $(n,m)$ particles A,B (probability $P(n,m)$) go into $(n+1,m)$ particles A,B (probability $P(n+1,m)$) with the total rate
$n\alpha_1$. For the total change  we get
\beq
\dot{P}(n,m)_{\alpha_1}=\alpha_1\Big[(n-1)P(n-1,m)-nP(n,m)\Big],\ \ n\geq 2.
\label{eq23}
\eeq

Transition $A\to B+B$. Now $(n+1,m-2)$ particles A,B (probability $P(n+1,m-2)$) go into $(n,m)$ particles A,B (probability $P(n.m)$)
with the total rate $\alpha_2(n+1)$ and $(n,m)$ particles (probability $P(n.m)$) go into $(n-1,m+2)$ particles (probability $P(n,m)$)
with the total rate $n\alpha_2$.
For the total change we get
\beq
\dot{P}(n,m)_{\alpha_2}=\alpha_2\Big[(n+1)P(n+1,m-2)-nP(n.m)\Big]. \ \ m\geq 2,
\label{eq24}
\eeq

Transition $B\to A+B$. In this case $(n-1,m)$ particles A,B (probability $P(n-1,m)$) go into $(n,m)$ particles (probability $P(n,m)$) with the
total rate $m\alpha_3$ and $(n,m)$ particles (probability $P(n,m)$)) go into $(n+1,m)$ particles (probability $P(n.m+1)$)
with the total rate $m\alpha_3$. For  the total change we get
\beq
\dot{P}(n,m)_{\alpha_3}=\alpha_3\Big[mP(n-1,m)-mP(n.m)\Big].\ \ n\geq 1.
\label{eq25}
\eeq

Transtion $A+A\to A$. In this case $(n+1)n/2$ pairs AA from $(n+1,m)$ particles (probability $P(n+1,m)$) transform into $(n,m)$ particles
(probability $P(n,m)$) with the total rate $(n+1)n\beta_1$ and $n(n-1)/2$ pairs from $(n,m)$ particles (probability $P(n,m)$) transform
into $(n-1,m)$ particles (probabilty $P(n-1,m)$) with the total rate $n(n-1)\beta_1$. For the total change we get
\beq
\dot{P}(n,m)_{\beta_1}=\beta_1\Big[(n+1)nP(n+1,m)-n(n-1)P(n,m)\Big].
\label{eq26}
\eeq

Transition $B+B\to A$. In this case $(m+2)(m+1)/2$ pairs BB from $(n-1,m+2)$ particles (probability $P(n-1,m+2)$) transform into $(n,m)$ particles
(probability $P(n,m)$) with the total rate $(m+2)(m+1)\beta_2$ and $m(m-1)/2$ pairs BB from $(n,m)$ particles
 transform into $(n+1,m-2)$ particles with the total rate
$m(m-1)\beta_2$. For the total change we find
\beq
\dot{P}(n,m)_{\beta_2}=\beta_2\Big[(m+2)(m+1)P(n-1,m+2)-m(m-1)P(n,m)\Big],\ \ n\geq 1.
\label{eq27}
\eeq

Transition $A+B\to B$. In this case
 $(n+1)m$ pairs AB from $(n+1,m)$ particles (probability $P(n+1,m)$) transform into $(n,m)$ particles
(probability $P(n,m)$) with the total rate $(n+1)m\beta_3$  and $nm$ pairs AB from $(n,m)$ particles (probabilty $P(n,m)$)
transform into $(n-1,m)$ particles (probability $P(n,m)$) with the total rate $nm\beta_3$. For the total change we get
\beq
\dot{P}(n,m)_{\beta_3}=\beta_3\Big[(n+1)mP(n+1,m)-nmP(n,m)\Big].
\label{eq27c}
\eeq

Inclusion of quartic couplings in the field-theoretical approach corresponds to inclusion of two more transitions
in the probabiistic framework
\[ A+A\to B+B\ \ (2\alpha_4),\ \ B+B\to A+A\ \ (2\beta_4).\]
The changes of the probability  are then as follows.

Process A+A$\to$B+B. (n+1)(n+2)/2 pairs AA from (n+2,m-2) particles (probabilitiy $P(n+2,m-2)$) transform into $(n,m)$ particles
(probability $P(n,m)$) with the total rate $(n+1)(n+2)\alpha_4$ and $n(n-1)/2$ pairs AA from $(n,m)$ particles
(probability $P(n,m)$)) transform into $(n-2,m+2)$ particles (probability $P(n-2,m+2)$) with rate $n(n-1)\alpha_4$.
For the total change we find
\beq
\dot{P}(n,m)_{\alpha_4}=\alpha_4\Big[(n+1)(n+2)P(n+2,m-2)-n(n-1)P(n.m)\Big],\ \ m\geq 2.
\label{eq27a}\eeq
In the process B+B$\to$ A+A particles A and B are interchanged. So we immediately find
\beq
\dot{P}(n,m)_{\beta_4}=\beta_4\Big[(m+1)(m+2)P(n-2,m+2)-m(m-1)P(n.m)\Big], \ \ n\geq 2.
\label{eq27b}\eeq

Note that restriction on the values of $n$ and $m$ are automatically taken into account assuming
that at unphyisical $n=-1,-2$ and $m=-1,-2$ the probabilities are zero.
\[P(n,-1)=P(n,-2)=P(-1,m)=P(-2,m)=0.
\]
The total evolution in rapidity is given by the sum
\beq
\dot{P}(n,m)=\sum_{i=1}^4\Big(\dot{P}(n,m)_{\alpha_i}+\dot{P}(n,m)_{\beta_i}\Big).
\label{eq29}
\eeq
One easily finds that in fact it is conserved in the evolution: $\dot{P}(n,m)=0$ (see Appendix 1.), so that one can put
\beq
\sum_{n,m}P(y,n,m)=1
\label {eq211}
\eeq
Also one can see that if at $y=0$ all probabilities are positive, they will remain such during evolution.
In fact assume that at a certain $y$ one of the probabilities $P(y,n,m)$ is equal to zero. However from the
evolution equation for $P(y,n.m)$ it follows that at this rapidity $dP(y,n,m)/dy$ is positive provided all $\alpha_i$ and $\beta_i$ are positive.
Then $P(y,n.m)$ cannot become negative at greater $y$ and remains positive at higher $y$.
As a result $P(y,n.m)$ satisfy all necessary conditions to be fulfilled for probabilities.

\subsection{Multiple moments and their evolution}
Multiple moments of particle distribution are defined in an obvious generalization of the pure pomeron case.
\beq
\nu(k,l)=\sum_{n\geq k.m\geq l}P(n,m)n(n-1)...(n-k+1)m(m-1)...(m-l+1).
\label{eq31}
\eeq
Since $n,m=0,1,2...$ with exception $n=m=0$, $\nu(k.l)$ are defined in a similar domain $k,l=0,1,...$ except $k=l=0$.
Their remarkable property is that their evolution in rapidity obeys simple equations expressing $d\nu/dy$ via $\nu$ themselves.
The evolution equations for $\nu(n,m)$ can be derived from their definition (\ref{eq31}) using the evolution equations for $P(n,m)$.
Siimilar to $P(n,m)$ they  split into 8 terms according to the 8 elementary transitions
\beq
\dot{\nu}(k,l)=\sum_{i=1}^4\Big(\dot{\nu}(k,l)_{\alpha_i}+\dot{\nu}(k,l)_{\beta_i}\Big).
\label{eq321}
\eeq
Their derivation is illustrated  in Appendix 2. As a result we find
\beq
\dot{\nu}(k,l)_{\alpha_1}=\alpha_1\Big[k\nu(k,l)+k(k-1)\nu(k-1,l)\Big].
\label{eq322}
\eeq
\beq
\dot{\nu}(k,l)_{\alpha_2}=\alpha_2\Big[-k\nu(k,l)+2l\nu(k+1,l-1)+l(l-1)\nu(k+1,l-2)\Big].
\label{eq323}\eeq
\beq
\dot{\nu}(k,l)_{\alpha_3}=\alpha_3\Big[kl\nu(k-1,l)+k\nu(k-1,l+1)\Big].
\label{eq323a}\eeq
\beq
\dot{\nu}(k,l)_{\beta_1}=
-\beta_1\Big[k\nu(k+1),l)+k(k-1)\nu(k.l)\Big].
\label{eq324}\eeq
\beq
\dot{\nu}(k,l)_{\beta_2}=\beta_2\Big[-l(l-1)\nu(k,l)-2l\nu(k,l+1)+k\nu(k-1,l+2)\Big].
\label{eq3251}
\eeq
\beq
\dot{\nu}(k,l)_{\beta_3}=-\beta_3)\Big[kl\nu(k-1,l)+k\nu(k-1,l+1)\Big].
\label{eq326}\eeq
\beq
\dot{\nu}(k,l)_{\alpha_4}=\alpha_4\Big[2l\nu(k+2,l-1)-2k\nu(k+1,l)+l(l-1)\nu(k+2,l-2)-k(k-1)\nu(k,l)\Big].
\label{eq327}\eeq
\beq
\dot{\nu}(k,l)_{\beta_4}=\beta_4\Big[2k\nu(k-1,l+2)-2l\nu(k,l+1)+k(k-1)\nu(k-2,l+-2)-l(l-1)\nu(k,l)\Big].
\label{eq328}\eeq

\section{From the probabilistic picture to the Quantum Field Theory}
In this section we constuct the
toy model which corresponds to the probabilistic picture assuming that the connection between them
is established by the relation which generalizes the case without odderonsin in the obvious manner.
Namely we present the wave function in the toy model as the Fock expansion in states $\Psi_{kl}$ of $k$ pomerons and $l$ odderons
\beq
\Psi(y)=\sum_{k,l}c(y,k,l)\Psi_{k.l}.
\label{cykl}\eeq
Here the summation goes over all $k,l\geq 0$ with $c(y,0,0)=0$ (in the following $y$ will be suppressed).
We connect the field-theoretical and probabilistic  pictures by the relarion betwee Fock's coefficients
in the former and averaged multiple moments in the latter generalizing (\ref{rel})
\beq
 c(k,l)=-(-\alpha)^k(-\beta)^l\frac{\nu(k,l)}{k!l!}
 \label{relg}
 \eeq
 with some constants $\alpha$ and $\beta$ determined by the adopted normlization.

 From the evolution equations for $\nu(k.l)$ derived in Section 2.2 we find the evolution equations for
 the Fock coefficients under 8 transitions with rates $\alpha_i$ and $\beta_i$ $i=1,2.3,4$:
\beq
\dot{c}(k,l)_{\alpha_1}
=\alpha_1\Big[kc(k.l)-\alpha(k-1)c(k-1,l)\Big],\label{eq44}\eeq
\beq
\dot{c}(k,l)_{\alpha_2}
=\alpha_2
\Big[-kc(k,l)+2\frac{\beta}{\alpha}(k+1)c(k=1,l-1)-\frac{\beta^2}{\alpha}(k+1)c(k+1,l-2)\Big],
\label{eq45}\eeq
\beq
\dot{c}(k,l)_{\alpha_3}
=\alpha_3\Big[-\alpha lc(k-1,l)+\frac{\alpha}{\beta}(l+1)c(k-1,l+1)\Big],
\label{eq46}\eeq
\beq
\dot{c}(k,l)_{\beta_1}
=-\beta_1\Big[-\frac{1}{\alpha}k(k+1)c(k+1,l)+k(k-1)c(k,l)\Big],
\label{eq47}\eeq
\beq
\dot{c}(k,l)_{\beta_2}
=-\beta_2\Big[l(l-1)c(k,l)-\frac{1}{\beta}l(l+1)c(k,l+1)+\frac{\alpha}{\beta^2}(l+1)(l+2)c(k-1,l+2)\Big],
\label{eq48}\eeq
\beq
\dot{c}(k,l)_{\beta_3}
=-\beta_3\Big[-\alpha lc(k-1,l)+\frac{\alpha}{\beta}(l+1)c(k-1,l+1)\Big].
\label{eq49}\eeq

\[\dot{c}(k,l)_{\alpha_4}=
\alpha_4\Big[-2\frac{\beta}{\alpha^2}(k+1)(k+2)c(k+2,l-1)+2\frac{1}{\alpha}k(k+1)c(k+1,l)\]\beq
+\frac{\beta^2}{\alpha^2}(k+1)(k+2)c(k+2,l-2)-k(k-1)c(k,l)\Big].\label{eqa2}\eeq

\[\dot{c}(k,l)_{\beta_4}=
\beta_4\Big[-2\frac{\alpha}{\beta^2}(l+1)(l+2)c(k-1,l+2)+2\frac{1}{\beta}l(l+1)c(k,l+1)\]\beq
+\frac{\alpha^2}{\beta^2}(l+1)(l+2)c(k-2,l+2)-l(l-1)c(k,l)\Big].\label{eqa2a}\eeq

To obtain the evolution equation for the wave function we use the $(u,w)$ representation
in which the creation operators for pomerons and odderons are mutiplcations by numbers $u$ and $w$ so that
the wave function is
\beq
\Psi(y,u,w)=\sum_{k,l}c(y,k,l)u^kw^l.\label{Psiuw}\eeq

Evolution equations for $\Psi$ are easily obtained from those for $c(k,l)$  (Appendix 3.).

 We sucsessively onsider the 6 transitions for triple interactions
\beq
\dot{\Psi}_{\alpha_1}=\alpha_1\Big(u\du-\alpha u^2\du\Big)\Psi,\label{dpa1}\eeq
\beq
\dot{\Psi}_{\alpha_2}=\alpha_2\Big(-u\du+2\frac{\beta}{\alpha}w\du-
\frac{\beta^2}{\alpha}w^2\du\Big)\Psi,\label{dpa2}\eeq
\[
\dot{\Psi}_{\alpha_3}=\alpha_3\Big(-\alpha uw\dw
+\frac{\alpha}{\beta}u\dw\Big) \Psi,\]
\beq
\dot{\Psi}_{\beta_1}=-\beta_1\Big(-\frac{1}{\alpha}u\ddu+u^2\ddu\Big)\Psi,\label{dpb1}\eeq
\[
\dot{\Psi}_{\beta_2}=-\beta_2\Big(w^2\ddw-\frac{1}{\beta} w\ddw +\frac{\alpha}{\beta^2}u\ddw\Big)\Psi,\]
\[
\dot{\Psi}_{\beta_3}=-\beta_3\Big(-\alpha uw\dw+\frac{\alpha}{\beta}u\dw\Big)\Psi.\]

The parts corresponding  to processes A+A$\leftrightarrow$B+B are
\beq
\dot{\Psi}_{\alpha_4}=-\alpha_4\Big(-2\frac{\beta}{\alpha^2}w+2\frac{1}{\alpha}u+\frac{\beta^2}{\alpha^2}w^2-u^2\Big)\frac{\pd^2}{\pd u^2}\Psi,
\label{dpa4}\eeq
\[
\dot{\Psi}_{\beta_4}=-\beta_4\Big(-2\frac{\alpha}{\beta^2}u+2\frac{1}{\beta}w+\frac{\alpha^2}{\beta^2}u^2-w^2\Big)\frac{\pd^2}{\pd w^2}\Psi.\]

As a result the total Hamiltonian corresponding to the probabilistic picture is a sum of 8 terms
\[
H=
\alpha_1\Big(-u\du+\alpha u^2\du\Big)+\beta_1\Big(-\frac{1}{\alpha}u\ddu+u^2\ddu\Big)\]\[
+\alpha_2\Big(u\du-2\frac{\beta}{\alpha}w\du+
\frac{\beta^2}{\alpha}w^2\du\Big)+\beta_2\Big(w^2\ddw-\frac{1}{\beta} w\ddw +\frac{\alpha}{\beta^2}u\ddw\Big)\]\[
+\alpha_3\Big(\alpha uw\dw
-\frac{\alpha}{\beta}u\dw\Big)+\beta_3\Big(-\alpha uw\dw+\frac{\alpha}{\beta}u\dw\Big)\]\beq
+\alpha_4\Big(-2\frac{\beta}{\alpha^2}w+2\frac{1}{\alpha}u+\frac{\beta^2}{\alpha^2}w^2-u^2\Big)\frac{\pd^2}{\pd u^2}
+\beta_4\Big(-2\frac{\alpha}{\beta^2}u+2\frac{1}{\beta}w+\frac{\alpha^2}{\beta^2}u^2-w^2\Big)\frac{\pd^2}{\pd w^2}.
\label{eq56}\eeq
Here the first line corresponds to the theory without odderons,
 the second and third  correspond to pomerons plus odderons
 with triple interactions, the fourth to inclusion of  quartic interactions.

In this theory without odderons one observes a relation between the coupling constants
which selects a very particular model to possess a probabilistic interpretation, the fact noticed in ~\cite{redif}. Indeed let the coupling constants be
\[\lambda_2=-\alpha_1,\ \lambda_3=\alpha_1\alpha,\ \ \lambda'_3=-\frac{\beta_1}{\alpha},\ \ \lambda_4=\beta_1\]
where $\lambda_{2,3,4}$ refers to $2,3,4$-fold interaction.  Obviously one finds a relation
\beq
\frac{\lambda_3}{\lambda_2}=\frac{\lambda_4}{\lambda'_3}=-\alpha,\label{condlam}\eeq
so that the four coupling constants cannot be taken arbirarily.
Under  condition (\ref{condlam}) the theory without odderons can indeed be considered as a probabilistic model.

The situation becomes worse and in fact disasrous after inclusion of odderons  The new terms in the Hamiltonian
violate the $C$ -nvariance and even admit direct transformtions of pomerons into odderons (second term  with $\alpha_2$).
Actually in all terms with odderons we find some with odd number of odderon variables, which is prohibited by $C$-parity conservation.

So unfortunately our conclusion is that in the presence of odderons and standard transformation laws under evolution one cannot
interprete the theory in  the probabilistic approach, at least in the technique using a particular relation between the
multiple moments and Fock's coefficients, whih was found working in the case without odderons..

\section{The Generalized Toy Regge-Gribov Model (GTRGM) with odderons.}
The minimal Toy model with odderons (witout quartic interactions) was proposed in  ~\cite{bkv2021}.
Here we present a slightly generalised Toy-Regge-Gribov model (GTRGM) with quartic interactions and general coupling constants.
As always the state is characterized by the wave function $\Psi(y)$ depending on rapidity $y$
and evolving according to the quasi-Schroedinger equation
\beq
\dot{\Psi}=-H\Psi.
\label{schr}
\eeq
The Hamiltonian depends on the pomeron creation and annihilation operators $\phi^\dagger$ and $\phi$ respectively and
on the odderon ones $\psi^\dagger$ and $\psi$:
\[
H=
= -\mu_P \phi^{\dagger}\phi - \mu_O \psi^{\dagger} \psi +
 i \lambda_3
\left( {\phi^\dagger}^2\phi + \phi^\dagger\phi^2\right)\]\[
+i2\lambda'_3\left(\psi^\dagger \psi \phi + \phi^\dagger \psi^\dagger \psi\right)+
i\bar{\lambda}_3\left( -\phi^{\dagger} \psi \psi + \psi^{\dagger} \psi^{\dagger} \phi \right)\]\beq
+\lambda_{41}{\phi^\dagger}^2\phi^2
+\lambda_{42} {\psi^\dagger}^2\psi^2+\lambda_{43} {\phi^\dagger}^2\psi^2\lambda_{44}{\psi^\dagger}^2\phi^2+
+\lambda_{45}\phi^\dagger\psi^\dagger\phi\psi.\label{hamg}
\eeq
This non-Hermitian Hamiltonian takes into account the symmetry under interchange target$\leftrightarrow$ projectile
and also
the negative signature of the odderon in the relative signs of the terms with $i\lambda_3$
From the analogy with the QCD framework it follows that all triple interaction constant are equal
\[\lambda_3=\lambda'_3=\bar{\lambda}=\lambda.\]
In the following we shall use this simplification.

To pass to real valuables one defines new operators $u$, $v$, $w$ and $z$ putting
\[ \phi=-iv,\ \ \phi^\dagger=-iu,\ \ \psi=-iz,\ \ \psi^\dagger=-iw\]
with  abnormal commutation relations
\[[v,u]=[z,w]=-1\]
and the vacuum state satisfying
\[v|0>=z|0>=0.\]
In terms of these new variables the Hamiltonian becomes real.

The scattering amplitude ${\cal A}$
is given by the matrix element
\beq
i{\cal A}(y)=<\Psi^{fin}|e^{-Hy}\Psi^{ini}>
\label{amp}\eeq
where $\Psi^{ini}$ and $\Psi^{fin}$ are the initial (at $y=0$) and final (at rapidity $y$) states.
Typically both are taken in the eikonal form
\begin{equation}
\Psi^{(ini)}=\Big(1-e^{-g_P^u- \sqrt{i}g_O w}\Big)|\Psi_0> ,\quad
\Psi^{(fin)}=\Big(1-e^{-g_P u+\sqrt{i}g_Ow}\Big)|\Psi_0>
\label{fin1}
\end{equation}
with couplings $g_P$ and $g_O$ of the pomeron and odderon to the two participants. (assumed equal for simplicity).
Note $\sqrt{i}$ and the difference in signs for the odderon coupling which again follow from its signature and makes the contribution
from the odderon exchange real.

Taking operators $u$ and $w$ as multiplcations ($u,w$-represetation) we find
\[ v=-\du,\ \ z=-\dw.\]
Then the Hamiltonial takes the form of a differental operator in variables $u$ and $w$:
\[H=-\mu_p u\du-\mu_Ow\dw
+
\lambda\Big(u^2\du-u\ddu+2wu\dw-2w\du\dw+u\ddw-w^2\du\Big)\]\beq
+\lambda_{41}u^2\ddu
+\lambda_{42}w^2\ddw+\lambda_{43} u^2\ddw+\lambda_{44}w^2\ddu+
+\lambda_{45}uw\du\dw\label{hamg1}.
\eeq

This Hamiltonian can be compared to the one following from the probabilistic model (\ref{eq56}).
In absence of odderons comparison of the double and triple interactions give
\[ \alpha_1=\mu_P,\ \ \alpha=\frac{\lambda}{\mu_P},\ \ \beta_1=\frac{\lambda^2}{\mu_p}\]
so that the quartic pomeron coupling is uniquely fixed  as $\beta_1$.
As mentioned this result was found in ~\cite{redif}:
the probabilstic interpretation is only possible for the Regge-Gribov model without odderons
only with a particular coupling for the quartic interaction.
 With odderons Hamiltonian (\ref{hamg1}) naturally conserves $C$-parity. It also contains the
 term correspomding to the odderon mass, absent in (\ref{eq56}). So the two Hamiltonians, the probabilstic and field theoretic,
  with odderons become fundamentally different.

It is of some interest to see to what sort of probabilities the GTRGM corresponds
if one derives then usung the same techniqe as in ~\cite{redif}, that is relating Fock's coefficients with multiple
moments. Obviously they have to be
probabilities which evolve in a differen=t manner as compared to those studied before, the latter possessing
 properties
naturally following from evolution of pomerons and odderons in the realistic world.

\section{Probabilities from GTRGM}
We present the wave function in the form of Fock's expanion in the states with a given number of pomerons and odderons (\ref{cykl}),
The Shroedinger equation (\ref{schr}) for the wave function generates evolution equations for the Fock coefficients $c(y.k,l)$
(partially derived in ~\cite{bkv2021} without quartic interactions). Taking the latter into account we find
\[
\dot{c}(k,l)=\mu_P kc(k,l)
-\lambda(k-1)c(k-1,l)+\lambda k(k+1)c(k+1,l)
\]
\[
+\mu_Olc(k,l)+2\lambda(k+1)lc(k+1,l)
-2\lambda lc(k-1,l)-\lambda (l+1)(l+2)c(k-1,l+2)\]\[
-\lambda (k+1)c(k+1,l-2)
-\lambda_{41} k(k-1)c(k,l)
-\lambda_{42}l(l-1)c(k,l),\]\beq
-\lambda_{43}(l+2)(l_1)c(k-2,l+2)
-\lambda_{44}(k+2)(k+1)c(k+2,l-2)
-\lambda_{45}kl c(k,l).
\label{fnm}
\end{equation}
where in the r.h.s $c(-1,l)=c(k,-1)=c(k,-2)=0$.
At $k=1,2$ amd $l=1,2$ terms $c(-2,l+2)$ and $c(k+2,-2)$ are actually absent.
So we assume $c(-2,l)=c(k,-2)=0$.
These equations will serve to introduce probabilities into GTRGM.
To do this we relate these
 Fock's coefficients
with the multiple average moments $\nu(k,l)$ in the probabilistic picture via  relation (\ref{relg})
with scaling coefficiens $a$ and $b$ dictated by the assumed normalization.
The evolution equations for Fock's coefficients in GTRGM then transform into evolution equation for $\nu(k.l)$
and aftterwards using the inverse to relation (\ref{eq31}) into evolution equations for probabilities $P(n.m)$, which dertermime them
once their values are given at zero rapidity.

We take  $a=\lambda/\mu_P$ from the comparison with the pure pomeron case in ~\cite{redif} and  $b=1$ for simplicity.
Denoting $\lambda'=\lambda^2/\mu_P$ from (\ref{relg}) we get for $\nu(k.l)$
\[
\dot{\nu}(k,l)=\mu_Pk\nu(k,l)+
\mu_Pk(k-1)\nu(k-1,l)-\lambda' k\nu(k+1,l)\]
\[+\mu_0\nu(k,l)-
2\lambda' l\nu(k+1,l)+2\mu_Pkl\nu(k-1,l)+
\mu_Pk\nu(k-1,l+2)\]
\[
+\lambda' l(l-1)\nu(k+1,l-2)
-\lambda_{41}k(k-1)\nu(k,l)
-\lambda_{42} l(l-1)\nu(k,l)
-\lambda_{43}\frac{\\mu_P^2}{\lambda^2}k(k-1)\nu(k-2,l+2),\]
\beq
-\lambda_{44}\frac{\lambda^2}{\mu_P^2}l(l-1)\nu_(k+2,l-2)
-\lambda_{45}kl\nu(k.l).\label{eq63}\eeq
where
\beq
\nu(-2,l)=\nu(k,-2)=\nu(y,-1,l)=\nu(y,k,-1)=\nu(y,k,-2)=\nu(y,0,0)=0.\label{condnu}\eeq

From these evolution equations we caculate the corresponding equations for probabilities using
thr formula inverse to (\ref{eq31})
\beq
 P(n,m)=\frac{T(n,m)}{n!m!}=\frac{1}{n!m!}\sum_{k=n,l=m}^\infty (-1)^{k-n}(-1)^{l-m}\nu(k,l)\frac{1}{(k-n)!(l-m)!}
 \label{eq64}\eeq
where $\nu(k.l)$ is expressed via $P(n,m)$ by (\ref{eq31}).
Here $n,m=0,1,...$ and $k,l=0,1,...$ except $n=m=k=l=0$/
Note that
Eqs, (\ref{eq31}) and (\ref{eq64}) connect $P(n,m)$ and $\nu(k,l)$
in the same domain of $(n,m)$ and $(k,l)$, namely $n,m,k,l=0,1,2...$ except $n=m=k=l=0$.

Calculatuion of the evolution equations for $T(n,m)=n!m!P(n.m)$ is presented in Appendix 4.

In the general case all 5 constants $\lambda_{4i}$, $i=1,...5$ can be arrbitrary subject only to the condition that
$V_4$ is posiive in the functional integral, which limits $\lambda_{41,42,45}$ to be positive and $|\lambda_{43,44}|$ to be less that
$\lambda_{41,42,45}$.

However we have seen that without odderons in the exceptional case when $\lambda_1=\lambda'$
the bad terms in the evolution equations are eliminated and the rest equations agree with the probabilistic structure.
Accordingly we also take $\lambda_1=\lambda'$ in the model with odderons,

Then our
evolution equations for $P(n,m)$ in GTRGM take the form.
\[
\dot{P}(n.m)=\mu_P\big[-nP(n,m)+(n-1)P(n-1,m)\Big]+\lambda'\Big[n(n+1)P(n+1,m)-n(n-1)P(n,m)\Big]\]\[
+\mu_O\Big[mP(n.m)-(m+1)P(n.m+1)\Big]
-2\lambda'(n+1)\Big[mP(n+1,m)-(m+1)P(n+1,m+1)\Big]\]\[
+2\mu_P\Big[mP(n-1,m)-mP(n,m)-(m+1)P(n-1,m+1)+(m+1)P(n,m+1)\Big]\]\[
+\mu_P(m+1)(m+2)\Big[P(n-1,m+2)-P(n,m+2)\Big]\]\[
+\lambda'(n+1)\Big[P(n+1,m-2)-2P(n+1,m-1)+P(n+1,m)\Big]\]\[
-\lambda_{42}\Big[m(m-1)P_(n,m)-2m(m=1)P(n,m+1)+(m+1)(m+2)P(n,m+2)\Big]\]\[
-\lambda_{43}\frac{\mu_P^2}{\lambda^2}(m+1)(m+2)\Big[ P(n-2,m+2)-2P(n-1,m+2)+P(n,m+2)\Big]\]\[
-\lambda_{44}\frac{\lambda^2}{\mu_P^2}(n+1)(n+2(\Big[P(n+2,m-2)-2P(n+2,m-1)+P(n+2,m)\Big]\]\beq
-\lambda_{45}\Big[nmP(n.m)-(n+1)mP(n+1,m)-n(m+1)P(n,m+1)+(n+1)(m+1)P(m+1,m+1)\Big].
\label{eq75}\eeq
where constants $\lambda_{42,43,44,45}$ related to odderons remain arbitrary.\\
We recall that $P(-1,m)=P(-2,m)=P(n,-1)=P(n-2)=0$

To compare we reproduce the most general probabilistic equations (\ref{eq29}) from Section 2:
\[
\dot{P}(n,m)=\alpha_1\Big[(n-1)P(n-1,m)-nP(n,m)\Big]+\beta_1\Big[(n+1)nP(n+1,m)-n(n-1)P(n,m)\Big]\]\[
+\alpha_2\Big[(n+1)P(n+1,m-2)-nP(n.m)\Big]+\beta_2\Big[(m+2)(m+1)P(n-1,m+2)-m(m-1)P(n,m)\Big]\]\[
+\alpha_3\Big[mP(n-1,m)-mP(n.m)\Big]+\beta_3\Big[(n+1)mP(n+1,m)-nmP(n,m)\Big]\]\[
+\alpha_4\Big[(n+1)(n+2)P(n+2,m-2)-n(n-1)P(n.m)\Big]\]\beq+\beta_4\Big[(m+1)(m+2)P(n-2,m+2)-m(m-1)P(n.m)\Big].
\label{pprob}\eeq

One observes that in the equations for the probabilities derived from GTRGM
the first lines have the correct form corresponsing to the probabilistic
interpretation  of the TRGM without odderons. So the problem is with  the rest terms  in (\ref{eq75}),
which appear after inclusion of odderons.

In (\ref{eq75}) their appear  seven  "correct" probabilities< which were  present in the probabilistic picture
(\ref{pprob})
\[P(n.m),\ \ P(n=1,m),\ \
P(n+1,m-2),\ \
P(n-1,m+2),\]\[
P(n+1,m)(n+1),\ \
P(n+2,m-2),\ \
 P(n-2,m+2),\]
although  the first two  enter with wrong coefficies as compared to (\ref{pprob}).

However (\ref{eq75})  also contains "incorrect" propabilities absent in the probabilistic picture.
They describe transitions  which contradict this picture\\
\[P(n+2,m)\ \ {\rm corresponds\ to}\ \ A+A\to vacuum,\]
\[P(n,m+2)\ \ {\rm corresponds\ to}\ \ B+B\to vacuum,\]
\[P(n-2,m)\ \ {\rm corresponds\ to}\ \ vacuum \to A+A,\]
\[P(n,m+1)\ \ {\rm corresponds\ to}\ \ B+B\to B.\]
\[P(n+1,m+1)\ \ {\rm corresponds\ to}\ \ A+A+B+B\to A+B,\]
\[P(n-1,m+1)\ \ {\rm corresponds\ to}\ \ A+B+B\to A+A+B,\]
\[P(n+1,m-1)\ \ {\rm corresponds\ to}\ \ A+A+B\to A+B+B,\]
\[P(n+2,m-1)\ \ {\rm corresponds\ to}\ \ A+A+B\to B+B.\]
These transitions are actually prohibited,
since they either violate $C$ -nvariance when the number of odderons is cnanged by unity, or
involve annihilation or
creation of particles from the vacuum.
So the probabilities generated in GTRGM by the techique developed in ~\cite{redif}, which relates Fock's coefficiens to multile moments,
turns out to be incompaticle with phyisical requirements.

\section{Probabilities from rescaled $c(k,l)$}
As discussed in the previous sections the method to introduce probabilities employed in
~\cite {redif} and based on the transition
of the Fock coefficients $c(k,l)$ according to the recipe
\beq
c(k,l)\to\nu(k.l)\to P(n,m)\label{recipe}\eeq
does not work for the models with odderons.   This gives motivation to search for alternative ways to introduce  probabilities in  the theory.
A simple method to achieve this goal was proposed in ~\cite{braun2025} for the model without odderons.
Then the Fock expansion of the wave function is
\[\Psi(y,u)=\sum_{n=1}c(y,k)u^k.\]
One may relate the probabilities directly to $c(k)$ avoiding their interptetation via averaged momenta $\nu(y,k)$.
In the "normal" theory with a real time one could normalize $\Psi$ to unity and then $|c(y,k)|^2$ would give the
probabilty to find $k$ pomerons in the system. In our case this is impossible, since the norm of the wave function
is not positive and depends on rapidity. However taking $|c(y,k)|^2$ as a relative measure to find $k$ pomerons one can rescale it
to achieve the desired normalisation and define the probabiliies as
\beq P(y,n)=|c(y,n)|^2R^{-k}(y)\label{pcp}\eeq
choosing $R(y)$ from the normalization condition
\[N(y,R)=\sum_{n=1}P(y,n)=1.\]
This procedure was realized in ~\cite{braun2025} and gave reasonable probabilties with the resuting entropy which did not grow
with rapidity so violently as the one  introduced accoerding to  (\ref{recipe}).

One is tempted to apply this rescaling method to the theory with odderons, since it guarantees construction of probabilities
satisfying all necessary conditions. However from the start one meets with the problem of generalization of the scaling (\ref{pcp})
to the case of two different particles. Parameter $R$ in (\ref{pcp}) had a meaning of the magnitude of the contribution from the pomeron.
Following this argument for the pomeron+odderon system one can take
\beq
P(y,n.m)=|C(y,n.m)|^2P_R^{-n}(y)R_O^{-m}(y).\label{pcpo}\eeq
with generally different scale factors $R_P(y)$ and $R_O(y)$. Unfortunately we have only one normalization condition to fix both scale factors
\[N(y,R_P,R_O)=\sum_{n+m\geq 0}P(y,n,m)=1.\]
so that  definition (\ref{pcpo}) gives different probabilities which  depend on the relation between $R_P$ and $R_O$.
So this definition   is not unique.

We will consider two limiting cases

{\bf Version A}
\beq
P(y,n.m)=|c(y,n.m)|^2R^{-(n+m)}(y).\label{pcpb}\eeq
 when the odderon gives the same growth as the pomeron, and

{\bf Version B}
\beq
P(y,n.m)=|c(y,n.m)|^2R^{-n}(y).\label{pcpa}\eeq
which one would naively expect when only the pomeron contributions grow,
since the "mass" of the odderon $\mu_O$ is standardly taken to be zero.

As our calculations demonstrate , there does not appear any qualitative difference beween these two versions.
However the quantitative difference is notable, which is illustrated in the following figures where the results are shown in both versions.

We apply this rescaling method to hA scattering, which in the quasi-classical approximation corresponds to fan diagrams.
However we shall also include triple interactions corresponding to fusion, dropping quartic coupling for simplicity. So our acting
Hamiltonian corresponds to (\ref{eq75}) with $\lambda'=\lambda_{4i}=0$, $i=2,3,4$.
Note that the direct evolution of $c(y,k,l)$ up to $y=30$ turned out to be impossiible due to
divergence at much smaller $y$. So we rather use the data from evolution of $P(y,n.m)$
derived from these $c(y,k.l)$ in the previous section and restore  from them the corresponding $c(y,k,l)$. In fact convergence of
 evolution of $P(y,n.m)$ is much better than of $c(y,k,l)$ and allows to achieve rapidities even substancially greater tan $y=30$.
We restricted the number of components $n,m$ by 80. The relative precision of our numerical results is estimated at the level
of $0.1\%$.

We consider both cases with $C$-parity equal 1
("pomeron") and -1 ("odderon") taking the initial wave function
\[\Psi(u,w)=u\ \ {\rm (pomeron)}\ \ {\rm and}\ \ \Psi(u,w)=w\ \ {\rm (odderon)}.\]
We evolved the wave function by the described simplified Hamiltonian up to rapidity $y=30$. Using (\ref{pcpa}) or (\ref{pcpb})
we then found  probabilties $P^{(P)}(n,m)$
and $P^{(O)}(n,m)$ for the $C=1$ and $C=-1$ cases respectively.

The overall probabilistic features of the model are characterized by the entropy
\beq
E(y)=-\sum_{n+m>0}P(y,n.m)\ln P(y,n,m).\label{ey}\eeq
The calculated entropy is shown in Fig. 1 for both cases $C=\pm 1$ and both variants A and B. For $C=+1$ ("pomeron") case also
the entropy in the model without odderon is shown.
\begin{figure}
\begin{center}
\epsfig{file=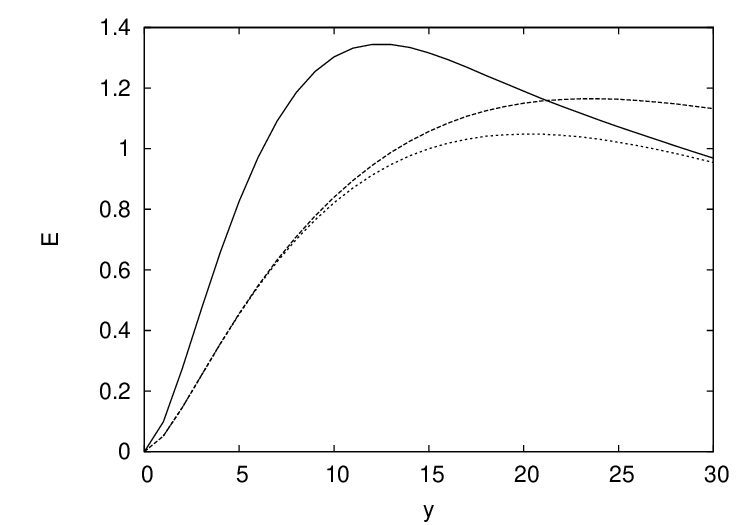, width=4 cm}
\epsfig{file=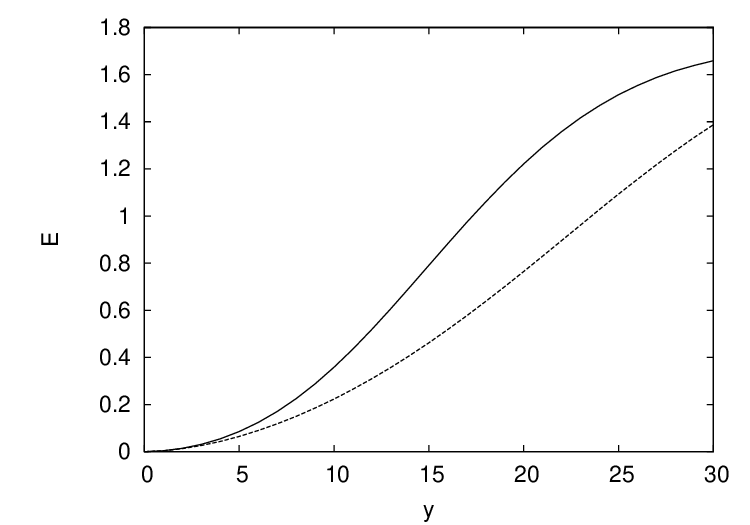, width=4 cm}
\caption{The entropy for the pomeron (left panel) and odderon (right panel) as a function of rapidity}
in versions A (solid lines) and B (dashed lines.
The entropy without odderons is shown by the lowest curve in the left panel)
\end{center}
\label{fig5r}
\end{figure}
As we observe in the pomeron sector the entropy is growing in all cases but slowlier than $y$ and moreover with
tendency to saturation or even diminishing (variant A) towards $y=30$. The difference between the
curves is notable but not very great, the entropy im variant A being somewhat
greater than in variant B (maximally by 30\% at $y=7$). Remarkably in variant B the influence of odderons is quite small.
In the odderon sector the entropy is of the same order as in the pomeron sector and grows in the similar manner.
Entropy  for variant A  is again greater than for variant B  to the same extent as for the pomeron sector.

More detailed information can be extracted from the probabilities themselves.
To present our results  more compactly we formed inclusive probablities to find for the pomeron case
$n$ pomerons $P^{(P)}_P(n)$ and
$m$ odderons $P^{(P)}_O(m)$
\[P^{(P)}_P(n)=\sum_mP^{(P)}(n,m),\ \ P^{(P)}_O(m)\sum_nP^{(P)}(n,m)\]
and similarly for the odderon case with $P^{(P)}\to P^{(O)}$.

For the initial pomeron our probabilities to find $n=1,5,10,15$ pomerons  after evolution
are shown in Fig. 2.
\begin{figure}
\begin{center}
\epsfig{file=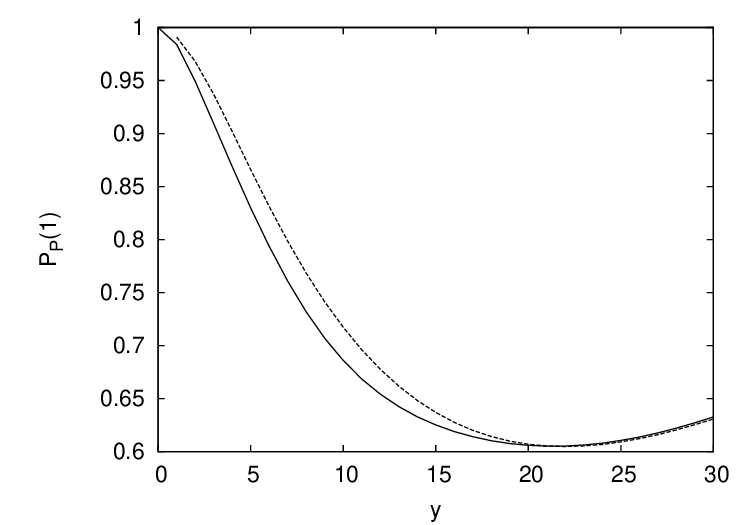, width=4 cm}
\epsfig{file=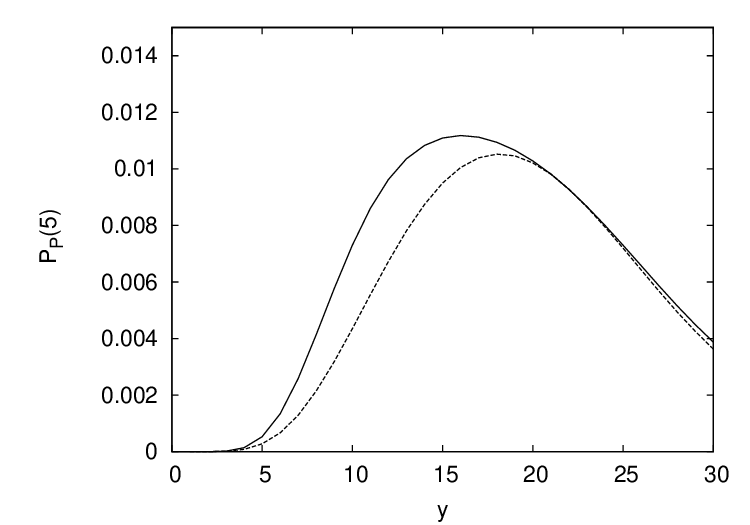, width=4 cm}
\epsfig{file=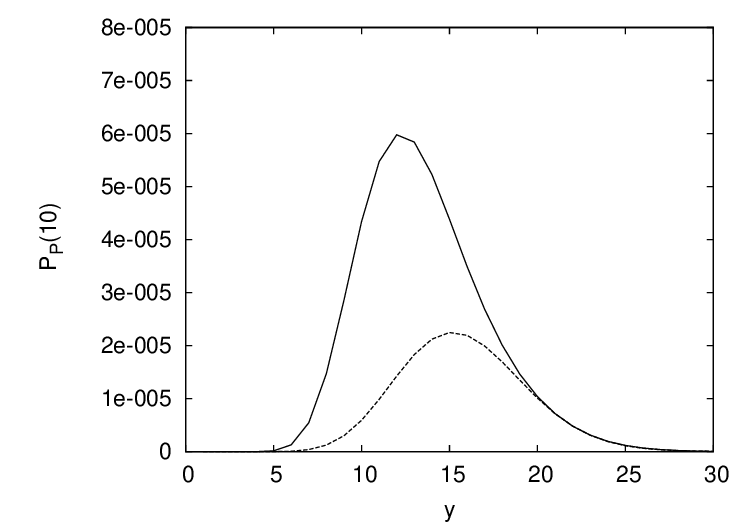, width=4 cm}
\epsfig{file=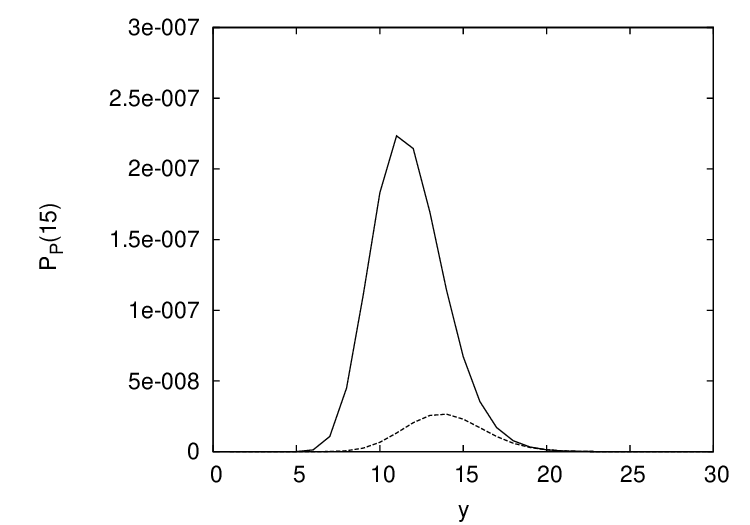, width=4 cm}
\caption{Inclusive probabilities $P^{(P)}_P(n)$ for $n=1,5,10,15$ at rapidities $y$
 in versiona A (solid lines) and B (dashed lines)}
\end{center}
\label{figs1a}
\end{figure}
For the same initial state the probabiliies to find $m=2,12$ odderons are shown in Fig. 3 in two left panels.
The probabilities to find other number of odderons are practically zero
\begin{figure}
\begin{center}
\epsfig{file=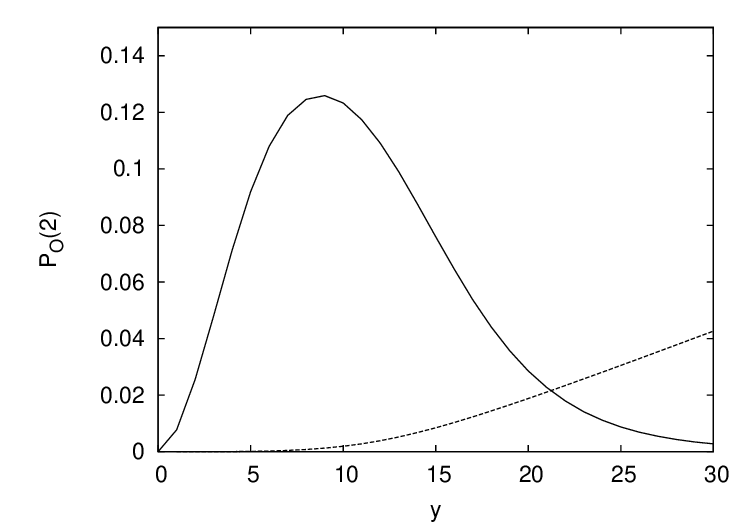, width=4 cm}
\epsfig{file=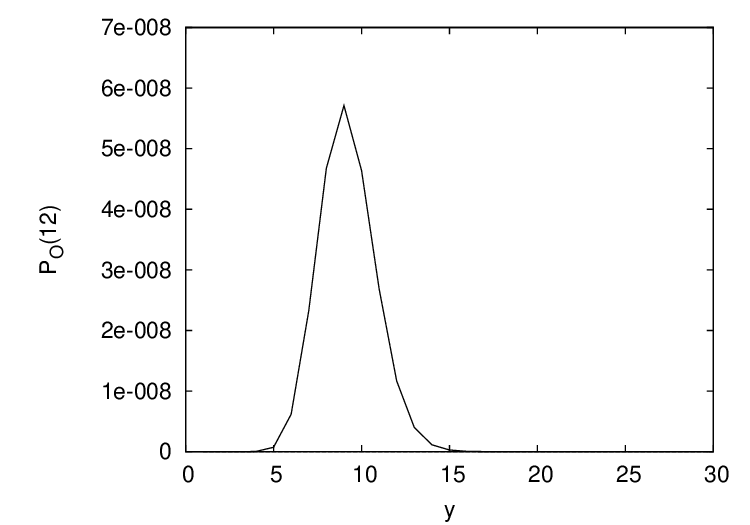, width=4 cm}
\epsfig{file=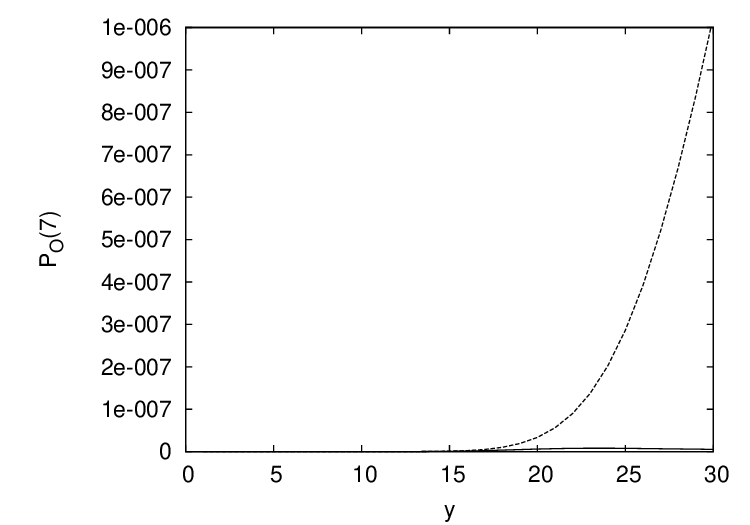, width=4 cm}
\caption{Inclusive probabilities $P^{(P)}_O(n)$ for $n=2,12$ (two left panels)
and $P^{(O)}_O(n=7)$ (right panel) at rapidities $y$.
in version A (solid lines) and B (dashed lines),
$P^{(O)}_O(1)=1$, at $n=2,12$ $P^{(O)}_O(n)=0$.
At $n=1,7$ the probabilities are zero}
\end{center}
\label{figs2a}
\end{figure}
The total probability to find pomerons is unity and  that of odderons is shown in Fig. 5
in the left panel.

For the initial odderon probabilities to find $n=1,5,10,15$ pomerons  after evolution
are shown in Fig. 4.
\begin{figure}
\begin{center}
\epsfig{file=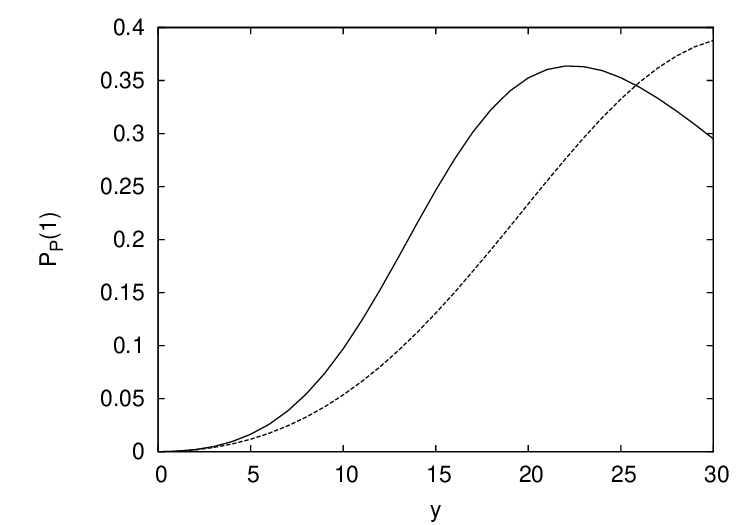, width=4 cm}
\epsfig{file=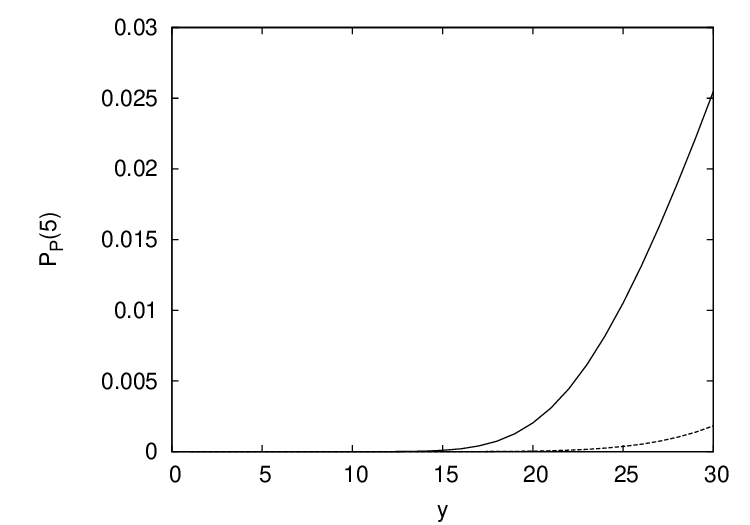, width=4 cm}
\epsfig{file=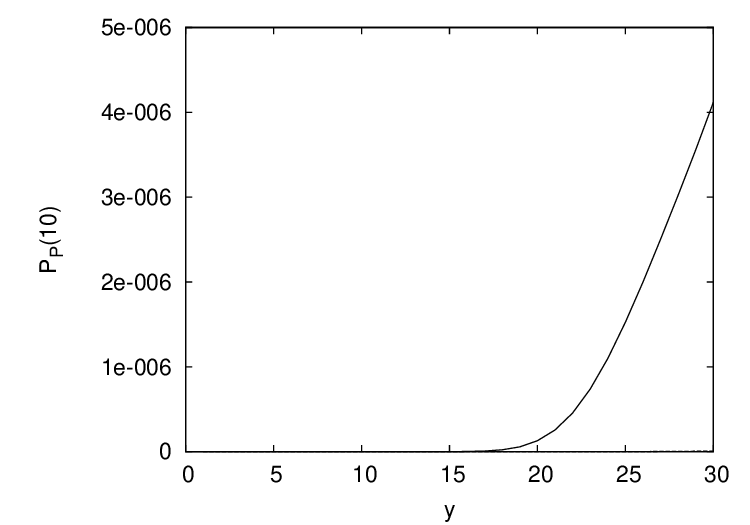, width=4 cm}
\epsfig{file=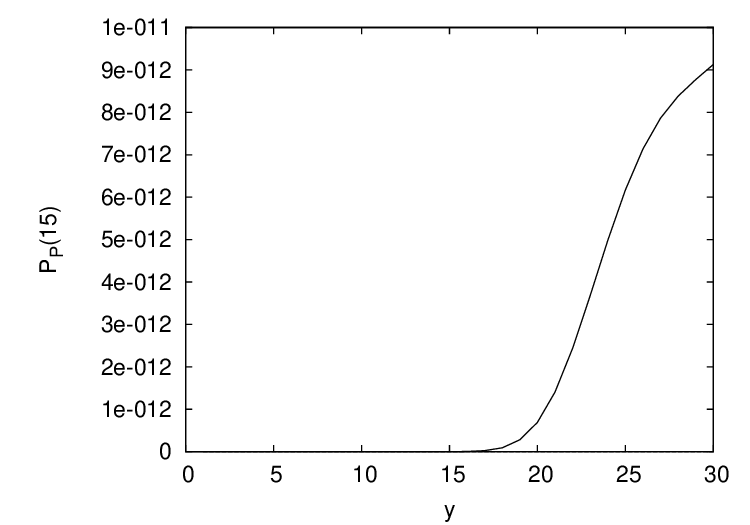, width=4 cm}
\caption{Inclusive probabilities $P^{(O)}_P(n)$ for $n=1,5,10,15$ at rapidities $y$
in versions A (solid lies) and B (dashed lines)}
\end{center}
\label{figs3a}
\end{figure}
The probability to find a single odderon in this case is unity, to find 7 odderons is shown Fig. 3 (rightmost panel)
and to find $m=2,12$  odderons is practically zero.

The total probability to find odderons is unity and  that of pomerons is shown in Fig. 5 in the right panel..
\begin{figure}
\begin{center}
\epsfig{file=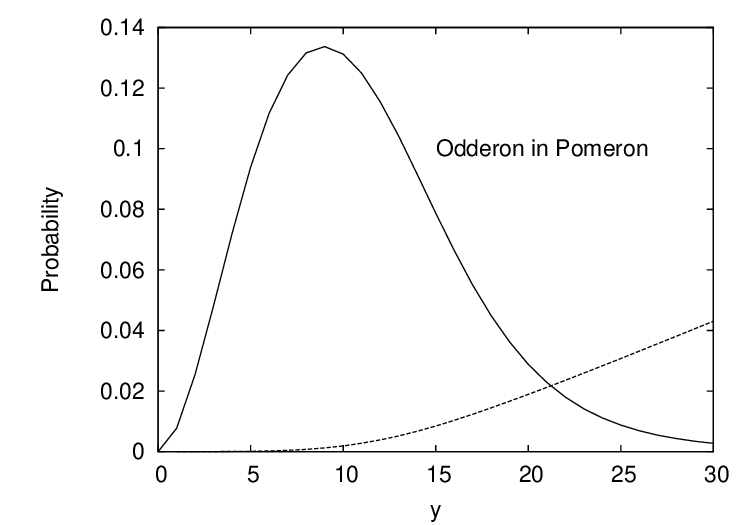, width=4 cm}
\epsfig{file=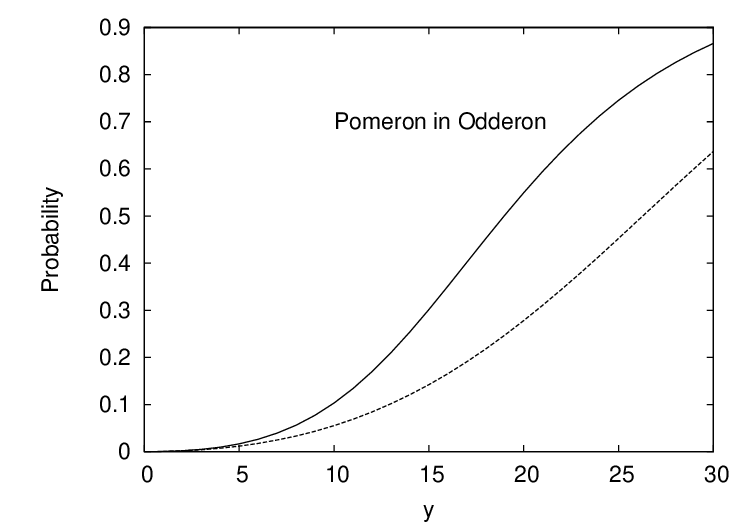, width=4 cm}
\caption{Probabilities to find any number of odderons in the pomeron sector (left panel)
and pomerons in the odderon sector (right panel) in versions A (solid lines) and B (dashed lines)}
\end{center}
\label{figs2b}
\end{figure}

Inspection of our numerical results leads to the following conclusions.

First of all both versions A and B give  similar entropies.
For the initial pomeron the entropy grows with rapidity until $y=20-25$ when it has a clear maximum (version A) or saturates (version B)
at the levek $E\sim 1.2$.
For the initial odderon in both versions the entropy is  growing with $y$ with a tendency to flattening towards y=30
when it reaches  practically the same level
$E=1.2-1.4$. So the main lesson from our method to introduce probabilities seems to be saturation of the entropy and even
going down at high eapidities,
in  contrast to lniear growt found in the probabilistic picture based on the identification $c\to\nu$. ~\cite{redif}.

Passing to concrete inclusive probabilities $P_P^{(P,O)}$ and  $P_O^{(P,O)}$ we observe that the calculated probabilties to find an odd
number of odderons in the pomeron and an even number of odderons in the odderon are all zero in agreement with the physical conservation laws.
This is a natural consequence of the  origin of our probabilities directly from the Fock coefficients, which of course obey this property.

One notes a remarkble behaviour of $P_P^{(P)}$ with rapidity to compare to  ~\cite{redif}. Namely instead of  flattening of the distribution
at large $y$ found there, in our rescaling method the probabilities at $n>1$ become peaked at around $y=10$ with the width of the peak
steadily contracting towards higher repidities.

Among other probabilities. apart from $P^{(O)}_O\simeq 1$ significant values are found for the production of a pair of odderons from the pomeron.
The form of its $y$-dependence is quite different for A and B versions. Wheras in B the probability steadily grows achieving value
$P^{(P)}_O(2)\simeq 0.04$ at $y=30$, in A it grows until $y=10$ up to $P^{(P)}_O(2)\simeq 0.12$
but then goes down to quite snall numbers at $y=30$

The rest probabiiites remain extremely small in the studied interval of rapidities, although they are rather different in versions A and B.

\section{Conclusions}
We have first investigated possibility to introduce  a probabilistic picture for the Toy model with pomerons and odderons
using the approach developed in ~\cite{redif}, which consists in relating the Fock coefficients in the development of the
wave function to  multiple moments of the probability distributions.
As shown in ~\cite{redif} in the model with only pomerons, this approach indeed allows to present as  a probabilistic theory
a particular  Toy model with the quartic interaction having an appropriately fixed coupling constant. In other versions, the one with
only triple interaction or with  a quartic interaction but with an arbitrary coupling constant
the probabilistic interpretation is impossible.
Our study shows that inclusion of odderons does not improve the situation. On the contrary, with odderons
the model cannot be transformed into a probabilistic one following the method of ~\cite{redif}.
Starting with the physically reasonable probabilities this method  leads to a completely unphysical field theory
which violates $C$ invariance. Inversely, starting from the physically admissible field model
one arrives at totaly  unphysical probabilities involving creation of particles from the vacuum or their annihilation.

As an alternative to introduce a probabilistic interpretation we used the approach developed in \cite{braun2025} to define
probabilities directly from Fock's coefficients suitably rescaling them to obey the standard requirements for probabilities.
This approach indeed allows to build probabilities for given numbers of pomerons and odderons. Physically they  manifesr themselves
in observed multiplicity distributions of hadrons generated by pomerons and odderons.
In case of two different particles, pomerons and odderons, our method is not uniquely fixed but depends on the relative
 magnitude of multiplicities coming from pomerons or odderons. We studied two limiting cases, with equal
 multiplicities and with the multiplicity from the odderon much smaller than that from the pomeron.
 The found entropies are  at most 30\% different at some particular rapidity.
 Remarkably  in the pomeron and odderon  sectors they turned out  to be quite similar both in magnitude and behavior in rapidity.
So in this respect one observes an approximate C independence. The probabilities themselves are found rapidly diminishing
with the number of particles. For different variants they are obtained significantly different only for large numbers of particles,
where they are  exceedingly small and hardly observable.

\section{Appendix 1, Conservation of probability}
Consider evolution of the total probability
\beq
N(y)=\sum_{n,m}P(y,n,m)
\label{eq210}
\eeq
for some particular transitions.
Starting with $\dot{N}_{\alpha_1}$ we have
\[\dot{N}_{\alpha_1}=\alpha_1\sum_{n,m}\Big((n-1)P(n-1,m)-nP(n,m)\Big)=
\alpha_1\Big(\sum_{n'=1,m=0}n'P(n',m)-\sum_{n=0,m=0}nP(n,m)\Big).\]
The contribution from $n=0$ in the second term is zero, so that we get $\dot{N}_{\alpha_1}=0$.

Next consider $\dot{N}_{\alpha_2}$:
\[\dot{N}_{\alpha_2}=\alpha_2\sum_{n,m}\Big((n+1)P(n+1,m-2)-nP(n,m)\Big)=
\alpha_2\Big(\sum_{n'=1,m'=0}n'P(n',m')-\sum_{n=0,m=0}nP(n,m)\Big).\]
Again the contribution from $n=0$ in the second term vanishes and we gee $\dot{N}_{\alpha_2}=0$.

Passing to $\beta$-evolution we have
\[\dot{N}_{\beta_1}=\beta_1\sum_{n,m}\Big((n+1)nP(n+1,m)-n(n-1)P(n,m)\Big)\]\[
=\beta_1\Big(\sum_{n'=1,m}n'(n'-1)P(n',m)-\sum_{n=0,m}n(n-1)P(n,m)\Big).\]
Contributions in both terms start with $n'=n=2$. So we gee $\dot{N}_{\beta_1}=0$.

Next $\dot{N}_{\beta_2}$.
\[\dot{N}_{\beta_2}=\beta_2\sum_{n,m}\Big((m+2)(m+1)P(n-1,m+2)-m(m-1)P(n,m)\Big)\]
\[=\beta_2\Big(\sum_{n'=0,m'=2}m'(m'-1)P(n',m')-\sum_{n=0.m=0}m(m-1)P(n,m)\Big).\]
In the second term only contributions with $m>1$ remain and we get $\dot{N}_{\beta_2}=0$.

In  absolutely the same way we find $\dot{N}_{\alpha_{3.4}}=\dot{N}_{\beta_{3.4}}=0$.
So in each of the 8  evolutions via $\alpha_i$ and $\beta_i$, $i=1,...4$ the total sum of probabilities is  separately
conserved and the total one can be put to unity.

|\section{Appendix 2. Evolution of  multiple moments}
We illustrate our technique on two examples, the simplest and most complicated cases.
The simplest case is transition A$\to$A+A. For this transition we  find
\beq
\dot{\nu}(k,l)_{\alpha_1}=\alpha_1\sum_{n,m}\Big[(n-1)P(n-1,m)-nP(n,m)\Big]n(n-1)...(n-k+1)m(m-1)...(m-l+1).
\label{eq33}
\eeq
Taking $n-1=n'$ we rewrite it as
\[
\dot{\nu}(k,l)_{\alpha_1}=\alpha_1\Big[
\sum_{n',m}n'P(n',m)(n'+1)n'(n'-1)...(n'-k+2)m(m-1)...(m-l+1)\]\[-
\sum_{n,m}nP(n',m)n(n'-1)...(n-k+1)m(m-1)...(m-l+1)\Big]\]\[=
\alpha_1\sum_{n,m}nP(n,m)\Big[(n+1)-(n-k+1)\Big]n(n-1)...(n-k+2)m(m-1)...(m-l+1)\]\beq=
\alpha_1\sum_{n,m}knP(n,m)n(n-1)...(n-k+2)m(m-1)...(m-l+1).
\label{eq34}
\eeq
In this case the odderon indices do not change and we suppress them, as well as the summation over $m$.
We introduce a notation
\[
[a(n)]=\sum_na(n)P(n)n(n-1)...(n-k+2),\]
so that we can write
$
\dot{\nu}(k,l)_{\alpha_1}=\alpha_1k[n]$.
To find ${n}$ we use two evident relations
$\nu(k)=[n-k+1]$ and $\nu(k-1)=[1]$.
from which it follows
$[n]=\nu(k)+(k-1)\nu(k-1)$
and so we get (\ref{eq322}).

We pass to the most complicated case of transitions A+A$\to$B+B.
Evolution equations for these transitions are (\ref{eq27a}) from which we get
\[\dot{\nu}(k,l)_{\alpha_4}=\alpha_4\sum_{n,m}n(n-1),,,(n-k+1)m(m-1)...(m-l+1)\]\[
\times\Big[(n+1)(n+2)P(n+2,m-2)-n(n-1)P(n,m)\Big]=t_1-t_2.\]
Here
\[t_1=\alpha_4\sum_{n,m}n(n-1),,,(n-k+1)m(m-1)...(m-l+1)(n+1)(n=2)P(n+2,m-2).\]
Or taking $n+2=n'$ and $m-2=m'$
\[t_1=\alpha_4\sum_{n',m'}n'(n'-1),,,(n'-k-1)(m'+2)(m'+1)m'(m-1)...(m-l+3)P(n',m')\]
\[t_2=\alpha_4\sum_{n,m}n^2(n-1^2(n-2)..(n-k+1)m(m-1)...(m-l+1)P(n.m).\]
So we find
\[\dot{\nu}(k,l)_{\alpha_4}=\alpha_4\sum_{n.m}P(n.m)n(n-1)...(n-k+1)m(m-1)...(m-l+3)\]\[
\Big[(n-k)(n-k-1)(m+2)(m+1)-n(n-1)(m-l+2)(m-l+1)\Big].\]
The square bracket is
\[\Big[...\Big]\equiv X=ln(n-1)(2m+3)-k(m+2)(m+1)(2n+1)-l^2n(n=1)+k^2(m+1)(m+2).\]

For further calculations we again introduce convenient notations. We suppress summation over $n$ and $m$ with weight
$\alpha_4P(n,m)$. We denote
\[n(n-1)...(n-k+1)=[k],\ \ m(m-1)...(m-l+3)=[l-2].\]
In these notations
\[\dot{\nu}(k,l)_{\alpha_4}=[k][l-2]X\equiv<X>.\]
Contributions to $<X>$ contain $<1>=c(k,l-2$ and 7 nontrivial terms $<x>$ where
\[x= n,\ m,\ nm,\ n^2,\ m^2,\ n^2m,\ m^2n.\]
In terms of these elementary averages we find
\[ <X>=2l<n^2m>-2k<nm^2>-l(l-3)<n^2>+k(k+1)<m^2>\]\[-(2l+6k)<nm>+(l^2-3l-4k)<n>+3k(k+1)<m>+2k(k+1)<1>.\]

Next using expressions for $\nu(k+\Delta_k.l-2+\Delta_l$ and choosing appropriate values for $\Delta_k$ and $\Delta_l$
we will express $<x>$ via $\nu$. For future calculations note
\[[k+1]=[k](n-k),\ \ [k+2]=[k](n-k)(n-k-1],\]
\[[l-1]=[l-2](m-l+2),\ \ [l]=[l-2]((m-l=2)(m-l+1).\]

We start from
\[\nu(k+1,l-2)=[k+1][l-2]=[k][l-2](n-k)=<n>-kc(k,l-2),\]
so
\[<n>=\nu(k+1,l-2)+k\nu(k,l-2).\]
Next
\[\nu(k,l-1)=[k][l-1]=[k][l-2](m-l+2)=<m>-(l-2)\nu(k,l-2),\]
so
\[<m>=\nu(k,l-1)+(l-2)\nu(k,l-2).\]
Now
\[\nu(k+1,l-1)=[k+1][l-1]=[k][l-2](n-k)(m-l+2)=[k][l-2]\Big(nm-(l-2)n-km+k(l-2)\Big)\]\[=<mn>-(l-2)<n>-k<m>+k(l-2)\nu(k,l-2)\]
which gives $<nm>$ once we know $<n>$ and $<m>$:
\[<nm>=\nu(k+1,l-1)-k(l-2)\nu(k,l-2)+(l-2)<n>+k<m>\]
Next
\[\nu(k+2,l-2)=[k+2][l-2]=[k][l-2](n-k)(n-k-1)\]\[=[k][l-2]\Big(n^2-n(2k+1)+k(k+1)\Big)=<n^2>-(2k+1)<n>+k(k+1)\nu(k.l-2).\]
from this
\[<n^2>=\nu(k+2,l-2)-k(k+1)\nu(k,l-2)+(2k+1)<n>.\]
Similarly
\[\nu(k,l)=[k][l]=[k][l-2](m-l+2)(m-l+1)=[k][l-2]\Big(m^2-m(2l-3)+(l-1)(l-2)\Big)\]
\[=<m^2>-(2l-3(<m>+(l-1)(l-2)\nu(k,l-2)\]
and so
\[<m^2>=\nu(k,l)-(l-1)(l-2)\nu(k,l-2)+(2l-3)<m>.\]
Now we consider
\[\nu(k+2,l-1)=[k+2][l-1]=[k][l-2](n-k)(n-k-1)(m-l+2)\]\[=
[k][l-2]\Big(n^2m-(2k+1)nm+k(k+1)m-(l-2)n^2+(l-2)(2k+1)n-(l-2)k(k+1)\Big)\]
This gives
\[<n^2m>=\nu(k+2,l-1)+(l-2)k(k+1)\nu(k,l-2)\]\[+(l-2)<n^2>+(2k+1)<nm>-(l-2)(2k+1)<n>-k(k+1)<m>\]
Finally
\[\nu(k+1,l)=[k+1][l]=[k][l-2](n-k)(m-l+2)(m-l+1)\]
\[=[k][l-2]
\Big(nm^2-nm(2l-3)+n(l-1)(l-2)
-km^2+k(2l-3)m-k(l-1)(l-2)\Big)\]
So we get
\[<nm^2>=\nu(k+1,l)+k(l-1)(l-2)\nu(k.l-2)+k<m^2>+(2l-3)<nm>\]\[-(l-1)(l-2)<n>-k(2l-3)<m>\]

Using these relations starting from $<n>$ and $<m>$ we successively calculate all the rest averages.
After simple but rather tedious calculations we finally find the desired evolution equation (\ref{eq327})
for transitions A+A$\to$B+B.

Applying this technique to the rest 7 transitions we arrive at the set of equations
(\ref{eq321})--(\ref{eq328}).

\section{Appendix 3. From $c(k,l)$ to $\Psi(u,w)$.}
 We consider some typical transitions

{\bf A$\to$A+A.}\\
\[\dot{\Psi}_{\alpha_1}=\sum u^kw^l\alpha_1\Big[kc(k,l)
-\alpha_1(k-1)c(k-1,l)\Big]=\alpha_1(t_1+t_2)\]
\[t_1=\sum u^kw^lkc(k,l)=\sum u\du u^kw^l c(r,l)=u\du\Psi\]
\[t_2=-\alpha\sum u^kw^l(k-1)c(k-1,l)=-\alpha\sum u^{k+1}w^lkc(k,l)\]\[=
-\alpha\sum u^2\du u^kw^lkc(k,l)=-\alpha u^2\du\Psi,\]
In the sum  we obtain (\ref{dpa1}).

{\bf A$\to$B+B.}\\
\[\dot{\Psi}_{\alpha_2}=\sum u^kw^l\alpha_2\Big[-kc(k,l)
+2\frac{\beta}{\alpha}
(k+1)c(k+1,l-1)-\frac{\beta^2}{\alpha}(k+1)c(k+1,l-2)\Big]=
\alpha_2(t_1+t_2+t_3).\]
\[t_1=\sum u^kw^l(-k)c(k,l)=\sum (-u\du)u^kw^lc(k,l)=-u\du\Psi.\]
\[t_2=2\frac{\beta}{\alpha}\sum u^kw^l(k+1)c(k+1,l-1)=
2\frac{\beta}{\alpha}\sum u^{k-1}w^{l+1}kc(k,l)\]\[=
2\frac{\beta}{\alpha}\sum w\du u^kw^{l}c(k,l)=2\frac{\beta}{\alpha}w\du\Psi.\]
\[t_3=-\frac{\beta^2}{\alpha}\sum u^kw^l(k+1)c(k+1,l-2)=
-\frac{\beta^2}{\alpha}\sum u^{k-1}w^{l+2}kc(k,l)\]\[=
-\frac{\beta^2}{\alpha}\sum w^2\du u^kw^lkc(k,l)=
-\frac{\beta^2}{\alpha}w^2\du\Psi.\]
In the sum we get (\ref{dpa2}).

{\bf A+A$\to$A.}\\
\[\dot{\Psi}_{\beta_1}=\sum u^kw^l(-\beta_1)\Big[-\frac{1}{\alpha}k(k+1)c(k+1,l)+k(k-1)c(k,l)\Big]=-\beta_1(t_1+t_2).\]
\[t_1=-\frac{1}{\alpha}\sum u^kw^lk(k+1)c(k+1,l)=-\frac{1}{\alpha}\sum k(k-1)u^{k-1}w^lc(k,l)\]\[=
-\frac{1}{\alpha}\sum u\ddu u^k w^lc(k,l)=-\frac{1}{\alpha} u\ddu\Psi.\]
\[t_2=\sum k(k-1u^kw^lc(k,l0\sum u^2\ddu u^kw^l c(k,l)=u^2\ddu\Psi.\]
Summing  we obtain (\ref{dpb1}).

{\bf A+A$\to$ B+B.}\\
We have for this transition evolution equation (\ref{eqa2})
We denote terms on the r.h.s as (1),...(4)
We find  for these terms  (suppressing $\alpha_4$)
\[(1)
=-2\frac{\beta}{\alpha^2}\sum(k+2)(k+1)c(k+2,l-1)u^kw^l
=-2\frac{\beta}{\alpha^2}\sum k(k-1)u^{k-2}w^{l+1}c(k,l)=-2\frac{\beta}{\alpha^2}w\frac{\pd^2}{\pd u^2}\Psi,\]
\[(2)=2\frac{1}{\alpha}\sum k(k+1)c(k+1,l)u^kw^l=2\frac{1}{\alpha}\sum k(k-1)u^{k-1}w^lc(k,l)=\frac{1}{\alpha}u\frac{\pd^2}{\pd u^2}\Psi,\]
\[(3)=\frac{\beta^2}{\alpha^2}\sum (k+2)(k+1)c(k+2,l-2)-\frac{\beta^2}{\alpha^2}\sum k(k-1)u^{k-2}w^{l+2}c(k,l)=
\frac{\beta^2}{\alpha^2}w^2\frac{\pd^2}{\pd u^2}\Psi.\]
\[(4)=-\sum k(k-1)c(k,l)u^kw^l=-u^2\frac{\pd^2}{\pd u^2}\Psi.\]
Summing these terms we get (\ref{dpa4}).

Parts of the Hamiltonian coming from the rest 4 transitions are calculated in the similar manner.

\section{Appendix 4. From Fock's coefficients to probabilities}
We will calculate $\dot{T}$ successively for terms 1 to 13 in (\ref{eq63}).
Terms 1,3,4  do not touch odderon variables, so we can suppress the latter and
pay attention to only pomerons. They belong to the minimal TRGM.

We start from  term 1, $\mu_pk(\nu(k)$-
\[\dot{T}_1(n)=\mu_P\sum_{k=n}(-1)^{k-n}k\frac{\nu(k)}{(k-n)!}=\sum_{k=n}(-1)^{k-n}\frac{1}{(k-n)!}\sum_{n'=k}\frac{T(n')}{(n'-k)!}=
\sum_{n'}Y^{(1)}_{nn'}T(n')\]
where
\[Y^{(1)}_{nn'}=\sum_k(-1)^{n-k}\frac{k}{(k-n)!(n'-k)!}.\]
Passing to $k'=k+n$
\[Y^{(1)}_{nn'}=\sum_k'(-1)^{k'}(k'+n)\frac{1}{k'!(n'-n-k')!}=nY_0(N)+Y_1(N),\ \ N=n'-n.\]
Here
\[Y_0(N)=\sum_k(-1)^k\frac{1}{k!((N-k)!}=\frac{1}{N!}\sum_k(-1)^kC_N^k=\delta_{N0},\]
\[Y_1(N)=\sum_k(-1)^k\frac{k}{k!(N-k)!}=\frac{1}{N!}\sum_{k}(-1)^k kC_N^k=-\delta_{N,1},\]
the last equality follows from the identities
\[\sum_k x^kC_N^k=(1+x)^N,\ \ \sum_kkx^kC)N^k=x\frac{d}{dx}(1+x)^N=Nx(1+x)^{N-1},\]\[
\sum_k k(k-1)x^kC_N^k=x^2\frac{d^2}{dx^2}(1+x)^N=N(N-1)x^2(1+x)^{N-2},\]
from which putting $x=-1$ we find
\[\sum_k(-1)^kC_N^k=\delta_{N0},\ \ \sum_k(-1)^kkC_N^k=\delta_{N0}=-\delta_{N1},\ \
\sum_k(-1)^kk(k-1)C_N^k=2\delta_{N2}\]
So after summation over $n'=n+N$ we get
\beq
\dot{T}_1(n)=\mu_p\Big[nT(n)-T(n+1)\Big].
\label{eq66}\eeq

Next
\[\dot{T}_3(n)=\mu_p\sum_{k=n}(-1)^{k-n}k(k-1)\frac{\nu(k-1)}{(k-n)!}\]\[=
\mu_p\sum_{k=n}(-1)^{k-n}\frac{1}{(k-n)!}\sum_{n'=k}\frac{T(n')}{(n'-k+1)!}
=\mu_P\sum_n'Y^{(3)}_{nn'}T(n'),\]
where
\[Y^{(3)}_{nn'}=\sum_k(-1)^{k-n}\frac{k(k-1)}{(k-n)!(n'-k+1)!}.\]

Passing to $k'=k+n$ we find factor
\[ (k'+n)(k'+n=1)=k'(k'+n-1)+n(k'+n-1)\]\[=k'(k'-1)+k'n+n(n-1)+k'n=n(n-1)+2k'n+k'(k'-1),\]
also we find in the denominator $(n'-k'-n+1)!=(N-k)!$ where now $N=n'-n+1$ so that $n'=n+N-1$.
So from these three terms we find
\[Y^{(3)}_{nn'}=n(n-1)Y_0-2nY_1+\frac{1}{2}Y_2=n(n-1)\delta_{n-1,n'}-2n\delta_{n'n}+\delta_{n+1,n'}\]
and so
\beq
\dot{T}_3(n)=\mu_P\Big[n(n-1)T(n-1)-2nT(n)+T(n+1)\Big].
\label{eq67}\eeq
In the sum we get
\[\dot{T}_{1+3}(n)=\mu_p\Big[nT(n)-T(n+1)+n(n-1)T(n-1)-2nT(n)+T(n+1)\Big]=\mu_P\Big[-nT(n)+n(n-1)T(n-1)\Big].\]
At $n=0$ the second term goes. As before this can be taken into account assuming $P(-1,m)=0$.

The 4-th term is the last one from TRGM:
\[
\dot{T}_4(n)=-\lambda'\sum_{k=n}(-1)^{k-n}\frac{k\nu(k+1)}{(k-n)!}=-\lambda'\sum_{k=n}(-1)^{k-n}\frac{k}{(k-n)!}
\sum_{n'=k}\frac{T(n')}{(n'-k-1)!}=-\lambda'\sum_{n'}Y^{(4)}_{nn'}T(n'),\]
where
\[
Y^{(4)}_{nn'}=\sum_k(-1)^{k-n}\frac{k}{(k-n)!(n'-k-1)!}=\sum_{k'}(-1)^{k'}\frac{k'+n}{k'!(n'-n-k-1)!}\]\[
=nY_0(N)+Y_1(N)=n\delta_{N0}-\delta_{N1}\]
and $N=n'-n-1$ so that $n'=n+N+1$.
Therefore
\[Y^{(4)}_{nn'}=n\delta_{n+1,n'}-\delta_{n+2,n'}\]
and
\[\dot{T}_4(n)=\lambda'\Big[T(n+2)-nT(n+1)\Big].\]

We pass to terms which depend explicitly on the odderon variables.
The second term in full analogy with the first one gives
\beq
\dot{T}_2(n,m)=\mu_0\Big[mT(n,m)-T(n,m+1)\Big]
\label{eq71}\eeq

Term No 5.
\[
\dot{T}_5(n,m)=-2\lambda\sum_{k=n,l=m}(-1)^{k-n}(-1)^{l-m}\frac{\nu(k+1,l)}{(k-n)!(l-n)!}\]\[
=-2\lambda\sum_{k=n}(-1)^{k-n}(-1)^{l-m}\frac{1}{(k-n)!(l-m)!}
\sum_{n'm'}\frac{T(n',m')}{(n'-k-1)!(m'-l)!}=\sum_{nn'}Y^{(5)}_{nn'}Y^{(5)}_{mm'}T(n',m')\]
where
\[Y^{(5)}_{nn'}=\sum_{k=n}(-1)^{k-n}\frac{1}{(k-n)!(n'-k-1)!}=\delta_{N0},\ \ N=n'-n-1.\]
\[Y^{(5)}_{mm'}=\sum_{l=m}(-1)^{l-m}\frac{l}{(l-m)!(m'-l)!}
=\sum_{l'}(-1)^{l'}\frac{m+l'}{l'!(M-l')!}=mY_0(M)+Y_1(M)=m\delta_{M0}-\delta_{M1}\]
where $M=m'-m$ and $m'=m+M'$.
So we find

\[\dot{T}_5(n,m)=-2\lambda' \Big[mT(n+1,m)-T(n+1,m+1)\Big]\]

Term No 6.
\[\dot{T}_6(n,m)=2\mu_p\sum_{k=n,l=m}(-1)^{k-n}(-1)^{l-m}\frac{kl \nu(k-1,l)}{(k-n)!(l-m)!}\]\[=
2\mu_p\sum_{k=n,l=m}(-1)^{k-n}(-1)^{l-m}\frac{kl}{(k-n)!(l-m)!}\sum_{n',m'}\frac{T(n',m')}{(n'-k+1)!)(m'-l)1}\]\[=
2\mu_P\sum_{n',m'}Y^{(6)}_{nn'}Y^{(6)}_{mm'}T(n',m')\]
where
\[Y^{(6)}_{nn'}=\sum_{k=n}(-1)^{k-n}\frac{k}{(k-n)!(n'-k+1)!},\ \
Y^{(6)}_{mm''}=\sum_{l=m}(-1)^{l-m}\frac{l}{(l-m)!(m'-l)!}\]
We introduce $k=k'+n$, $N=n'-n+1$ and $l=l'+m$, $M=m'-m$ to obtain
\[Y^{(6)}_{nn'}=\sum_{k'}(-1)^{k'}\frac{k'+n}{(k'!(M-k')!}=nY_0(N)+Y_1(N)=n\delta_{N0}-\delta_{N1},\]
\[Y^{(6)}_{mm''}=\sum_{l'}(-1)^{l'}\frac{l}{l'!(M-l')!}=mY_0{M}+Y_1(M)=m\delta_{M0}-\delta_{M1}.\]
This gives
\[\dot{T}_6(n,m)=2\mu_P\Big[nmT(k-1,m)-mT(n,m)-nT(n-1,m+1)+T(n,m+1)\Big]\]
Here $T(-1,m+1)=0$

Term No 7.
\[\dot{T}_7(n,m)=\mu_p\sum_{k=n,l'=m}(-1)^{k-n}(-1)^{l-m}\frac{k \nu(k-1,l+2)}{(k-n)!(l-m)!}\]
\[=
\mu_p\sum_{k=n,l=m}(-1)^{k-n}(-1)^{l-m}\frac{k}{(k-n)!(l-m)!}\sum_{n',m'}\frac{T(n',m')}{(n'-k+1)!(m'-l-2)!}\]
\[=
\mu_P\sum_{n',m'}Y^{(7)}_{nn'}Y^{(7)}_{mm'}T(n',m')\]
where
\[Y^{(7)}_{nn'}=\sum_{k=n}(-1)^{k-n}\frac{k}{(k-n)!(n'-k+1)!},\ \
Y^{(7)}_{mm'}=\sum_{l=m}(-1)^{l-m}\frac{l}{(l-m)!(m'-l-2)!}.\ \]
We introduce $k=k'+n$, $N=n'-n+1$, $n'=n+N-1$  and $l=l'+m$, $M=m'-m-2$, $m'=m+M+2$ to obtain
\[Y^{(7)}_{nn'}=\sum_{k'}(-1)^{k'}\frac{k'+n}{(k'!(M-k')!}=nY_0(N)+Y_1(N)=n\delta_{N0}-\delta_{N1}=n\delta_{n-1,n'}-\delta_{nn'},\]
\[Y^{(7)}_{mm''}=\sum_{l'}(-1)^{l'}\frac{1}{l'1!(M-l')!}=Y_0(M)=\delta_{m+2,m'}.\]
So we get
\[\dot{T}_7(n,m)=\mu_P\Big[nT(n-1,m+2)-T(n,m+2)\Big]\]
with $T(-1,m+2)=0$

Term No 8.
\[\dot{T}_8(n,m)=\lambda'\sum_{k=n,l=m}(-1)^{k-n}(-1)^{l-m}\frac{l(l-1) \nu(k+1,l-2)}{(k-n)!(l-m)!}\]
\[=\lambda'\sum_{k=n,l=m}(-1)^{k-n}(-1)^{l-m}\frac{l(l-1)}{(k-n)!(l-m)!}\sum_{n',m'}\frac{T(n',m')}{(n'-k-1)!(m'-l+2)1}\]\[=
\lambda'\sum_{n',m'}Y^{(8)}_{nn'}Y^{(8)}_{mm'}T(n',m')\]
where
\[Y^{(8)}_{nn'}=\sum_{k=n}(-1)^{k-n}\frac{1}{(k-n)!(n'-k-1)!},\ \
Y^{(8)}_{mm'}=\sum_{l=m}(-1)^{l-m}\frac{l(l-1)}{(l-m)!(m'-l+2)!}.\ \]
We introduce $k=k'+n$, $N=n'-n-1$ so that $n'=n+N+1$ and $l=l'+m$, $M=m'-m+2$, so that $m'=m+M-2$.
In the numerator we find
\[(l'+m)(l'+m-1)=l'(l'-1)+m(m-1)+2ml'.\]
As a result we find with $N=n'-n-1$ and $M=m'-m+2$
\[Y^{(8)}_{nn'}=\sum_{k'}(-1)^{k'}\frac{1}{k'!(N-k')!}=Y_0(N)=\delta_{N0}=\delta_{n_1,n'},\]
\[Y^{(8)}_{mm'}=\sum_{l'}(-1)^{l'}\frac{l'(l'-1)+m(m-1)+2l'm}{l'!(M-l')!}=m(m-1)Y_0(M)+2mY_1(M)+\frac{1}{2}Y_2(M)\]
\[=
m(m-1)\delta_{M0}-2m\delta_{M1}+\delta_{M2}=m(m-1)\delta_{m-2,m'}-2m\delta_{m-1,m'}+\delta_{mm'}.\]
This gives
\[\dot{T}_8(n,m)=\lambda'\Big[m(m-1)T(n+1,m-2)-2mT(n+1,m-1)+T(n+1,m)\Big]\]
With $m=0,1$ the first terms are absent, So we assume $T(n,-1)=T(n,-2)=0$

Now contributions from the quartic interactions.
\[
T^{(9)}(n,m)=\sum_{k=n,l=m}(-1)^{k+l-n-m}\frac{1}{(k-n)!(l-m)!}
\Big[-\lambda_1 k(k-1)\nu(k,l)\Big]
\]
\[=
-\lambda_1\sum_{k=n,k=m}(-1)^{k+l-n=m}\frac{k(k-1)}{(k-n)!(l-m)!}\sum_{n'=k,m'=l}\frac{T(n',m')}{(n'-k)!(m'-l)!}\]\[=
-\lambda_1\sum_{n',m'}T(n',m')Y^{(1)}_{nn'}Y^{(1)}_{mm'}.\]
Here
\[Y^{(9)}_{nn'}=\sum_{k=n}(-1)^{k-n}\frac{k(k-1)}{(k-n)!(n'-k)!}=\sum{k'=0}(-1)^{k'}\frac{(k'+n)(k'+n-1)}{k'!(N-k')!}\]
and $N=n'-n$. The numerator is
$ n(n-1+2kn+k(k-1)$,
so that we get
\[Y^{(9)}_{nn'}=n(n-1)\delta_{N0}-2n\delta_{N1}+k(k-1)\delta_{N2}.\]
A similar calculation  gives $Y^{(1)}_{mm'}=\delta_{mm'}$.
So we obtain
\[\dot{T}^{(9)}(n.m)=-\lambda_2\sum_{n'm'}\delta_{mm'}\Big(n(n-1)\delta_{nn'}-2n\delta_{n+1,n'}+\delta_{n+2,n'}\Big)\]
\[
-\lambda_1\Big[n(n-1T(n,m)-2nT(n+1,m)+T(n+2,m)\Big].\]

By symmetry we immediately get
\[
\dot{T}^{(10)}=-\lambda_2\Big[m(m-1)T(n,m)-2mT(n,m+1)+T(n,m+2)\Big].\]

We pass to $T^{(11)}$.
\[\dot{T}^{(11)}(n,m)=\sum_{k=n,l=m}(-1)^{k+l-n-m}\frac{1}{(k-n)!(l-m)!}\Big[-\lambda_3\frac{\beta^2}{\alpha^2}k(k-1)\nu(k-2,l+2)\Big]
\]
\[=\lambda_3\frac{\beta^2}{\alpha^2}\sum_{k=n,l=m}(-1)^{k+l-n-m}\frac{k(k-1)}{(k-n)!(l-m)!}
\sum_{n'=k-2,m'=l+2}\frac{T(n',m')}{(n'-k+2)!(m'-l-2)!}\]
\[
=-\lambda_3\sum_{n',m'}Y^{(3)}_{nn"}Y^{(3)}_{mm'}.\]
We have
\[Y^{(11)}_{nn'}=\sum_{k=n}(-1)^{k-n}\frac{k(k-1)}{(k-n)!(n'-k+2)!}
=\sum_{k'=0}(-1)^{k'}\frac{(k'+n)(k'+n-1)}{k'!(N-k')!}\]
where $N=n'-n+2$.
We find
\[Y^{(11)}_{nn'}=n(n-1)\delta_{N0}-2n\delta_{N1}+\delta_{N2}=n(n-1)\delta_{n-2,n'}-2n\delta_{n-1,n'}+\delta_{nn'}.\]
In the same manner we find
\[Y^{(11)}_{mm'}=\delta_{M0}=\delta_{m+2,m'},\ \ M=m'-m-2.\]
So in the end
\beq
\dot{T}^{(11)}(n,m)=-\lambda_3\frac{\beta^2}{\alpha^2}\Big[n(n-1)T(n-2,m+2)-2nT(n-1,m+2)+T(n,m+2)\Big]
\label{t3a}\eeq
and by symmetry
\beq
\dot{T}^{(12)}(n,m)=-\lambda_4\frac{\alpha^2}{\beta^2}\Big[m(m-1)T(n+2,m-2)-2mT(n+2,m-1)+T(n+2,m)\Big].
\label{t3b}\eeq
where we assume $T(-2,m)=T(n,-2)=0$.

Finally
\[\dot{T}^{(13)}(n,m)=\sum_{k=n,ml=m}(-1)^{k+l-n-m}\frac{1}{(k-n)!(l-m)!}\Big[-\lambda_5 klc(k,l)\Big]\]\[=
-\lambda_5\sum_{k=n,l=m}(-1)^{k+l-n-m}\frac{kl}{(k-n)!(l-m)!}\sum_{n',m'}\frac{T(n',m')}{(n'-k)!(m'-l)!}\]\[=
-\lambda_5\sum_{n',m'}T(n',m')Y^{(5)}_{nn'}Y^{(5)}_{mm'}\]
where
\[Y^{(13)}_{nn'}=\sum_{k=n}(-1)^{k-n}\frac{k}{(k-n)!(n'-k)!}=\sum_{k'=0}(-1)^{k'}\frac{k'+n}{k'!(N-k')!}
=n\delta_{N0}-\delta_{N1}\]
and $N=n'-n$. Matrix  $Y^{(5)}_{mm'}$ is the same.
So we get
\[\dot{T}^{(13)}=-\lambda_5\sum_{n',m'}T(n',m')\Big(n\delta_{nn'}-\delta_{n+1,n'}\Big)
\Big(m\delta_{mm'}-\delta_{m+1,m'}\Big)\]\[=
-\lambda_5\Big[nmT(n,m)-mT(n+1,m)-nT(n,m+1)+T(n+1,m+1)\Big].\]

\end{document}